\documentclass[aps,nofootinbib,preprintnumbers,nobalancelastpage]{revtex4}
\pdfoutput = 1
\usepackage[english]{babel}
\usepackage{amsmath,amssymb}
\usepackage[pdftex]{graphicx}
\usepackage[sort&compress]{natbib}
\usepackage{amsfonts}
\usepackage{bbm}
\usepackage{color}

\def\svann{{<\sigma v>_{ann}}}

\def\gsim{\mathrel{\lower3pt\hbox{$\sim$}}\hskip-11.5pt\raise3pt\hbox{$>$}\;}
\def\lsim{\mathrel{\lower3pt\hbox{$\sim$}}\hskip-11.5pt\raise3pt\hbox{$<$}\;}
\newcommand{\AddrULB}{%
 Service de Physique Th\'eorique, Universit\'e Libre de Bruxelles,\\
CP225, Bld du Triomphe, 1050 Brussels, Belgium}

\begin{document}

\preprint{ULB-TH/10-25}

\title{Constraints on light WIMP candidates from the Isotropic Diffuse Gamma-Ray Emission}

\vspace{3mm}
\author{Chiara Arina}\email{carina@ulb.ac.be}
\author{Michel H.G.~Tytgat}\email{mtytgat@ulb.ac.be}

\affiliation{\AddrULB}

\begin{abstract}

Motivated by the measurements reported by direct detection experiments, most notably DAMA, CDMS-II, CoGeNT and Xenon10/100, we study further the constraints that might be set on some light dark matter candidates, $M_{DM} \sim $ few GeV, using the Fermi-LAT data on the isotropic gamma-ray diffuse emission.  
In particular, we consider a Dirac fermion singlet interacting through a new $Z^\prime$ gauge boson, and a scalar singlet $S$ interacting through the Higgs portal. Both candidates are WIMP (Weakly Interacting Massive Particles), {\em i.e.} they have an annihilation cross-section in the pbarn range. Also they may both have a spin-independent elastic cross section on nucleons in the range required by direct detection experiments. Although being generic WIMP candidates, because they have different interactions with Standard Model particles, their phenomenology regarding the isotropic diffuse gamma-ray emission is quite distinct.  
In the case of the scalar singlet, the one-to-one correspondence between its annihilation cross-section and its spin-independent elastic scattering cross-section permits to express the constraints from the Fermi-LAT data in the direct detection exclusion plot, $\sigma_n^0-M_{DM}$.  Depending on the astrophysics, we argue that it is  possible to exclude the singlet scalar dark matter candidate at $95\%$ confidence level. The constraints on the Dirac singlet interacting through a $Z^\prime$ are comparatively weaker. 
\end{abstract}
\maketitle


\section{Introduction}

The cosmological data show evidence of dark matter (DM)~\cite{Komatsu:2010fb}, but the nature of the DM particle, if any, is still unknown. 
Recently, some interest has been taken in a light WIMP\footnote{By a WIMP {(Weakly Interacting Massive Particle)} we mean a cold dark matter candidate, with an annihilation cross-section of the order of 1 pbarn, so that its relic abundance may be fixed by the standard thermal freeze-out mechanism. This potentially includes many candidates, whose interactions might have nothing (or little) to do with weak interactions. By a light WIMP, we mean here a candidate in the mass range between 1 and 15 GeV, substantially lighter than the usual suspects, like the neutralino.}, with  $M_{DM} \lsim 15$ GeV. This is largely motivated by the modulation observed by the DAMA/LIBRA and DAMA/NaI experiments~\cite{Bernabei:2008yi}, the excess of events at low recoil energies measured by CoGeNT~\cite{Aalseth:2010vx}, and an anomaly reported by the CRESST collaboration at recent conferences~\cite{CRESST_collar}. If interpreted in terms of Spin Independent (SI) elastic scattering of a DM particle, these experiments all point to a candidate with a mass in the few GeV range, and a SI cross-section $\sigma_n^0 = {\cal O}(10^{-40})$ cm$^2$~\cite{Petriello:2008jj,Savage:2008er,Kopp:2009qt,Fitzpatrick:2010em,Andreas:2010dz,Chang:2010yk}. Whether these results, to which we will loosely refer as CoGeNT-DAMA, are due to dark matter or to a more mundane sort of background is a matter of debate (see e.g.~\cite{Kudryavtsev:2009gd,Ralston:2010bd,Bernabei:2010ke}). Furthermore other direct detection experiments have set exclusion limits that may exclude part or whole regions of the parameter space favoured by CoGeNT-DAMA (see for instance~\cite{Kopp:2009qt} and~\cite{Hooper:2010uy}). Most relevant for the low mass region, $M_{DM} \lsim 15$ GeV, are the CDMS-Si~\cite{Akerib:2005kh}, CDMS-II~\cite{Ahmed:2009zw},  Xenon10~\cite{Angle:2008we,Angle:2009xb} and Xenon100~\cite{Aprile:2010um} exclusion limits. 
However it is quite delicate to set robust exclusion limits in the low mass region, which in the direct detection experiments corresponds to small recoil energies\footnote{The exclusion limits may be very sensitive to parameters which are poorly constrained, see for instance the discussion revolving around the scintillation efficiency in LXe experiments~\cite{Collar:2010gg,Collaboration:2010er,Collar:2010gd,Andreas:2010dz,Savage:2010tg,Collar:2010nx}.}, and it may pay to look further for alternative ways to constrain a light WIMP. 

A most interesting possibility in the near future is  to look for missing energy at colliders, and it has been shown that, depending on the assumed properties of the light WIMP, the limits by the Tevatron and the LHC may be competitive with direct searches for light candidates, $M_{DM} \lsim 10$ GeV~\cite{Goodman:2010yf,Bai:2010hh}. The next best possibility is to consider indirect signatures that are specific to DM, {\em i.e.} with no or little background from known physics, like a monochromatic gamma-ray line on the sky, or a flux of energetic neutrinos from the Sun. The former signature depends very much on the dark matter candidate. The latter however is quite generic for a light WIMP relevant for CoGeNT-DAMA, at least because they  must have a large scattering cross-section on matter, and thus may be efficiently captured in the Sun. Their subsequent annihilation at the centre of the Sun would release a flux of neutrinos which depends on the annihilation channels of the DM candidate, and may be constrained by Super-Kamionkande~\cite{Savage:2008er,Feng:2008qn,Andreas:2009hj,Fitzpatrick:2010em,Niro:2009mw}. 

The other indirect signatures, be it a continuous gamma-ray spectrum or extra antimatter in cosmic rays,  are more prone to uncertainties, most notably because of large and not well-understood backgrounds from astrophysical sources. Nevertheless their measurement by the new generation of instruments (in particular Fermi-LAT) are expected to put very relevant constraints on a light WIMP. Indeed, although the annihilation of a light WIMP produces few particles (for instance, a 10 GeV candidate annihilating into $b\bar b$ quarks would give on average ${\cal O}(10)$ photons), this is more than compensated by the  $1/M_{DM}^2$ dependence of the fluxes, for fixed energy density $\rho_{DM}$. This is of course well-appreciated, and, incidentally, the current analysis already  give constraints on the annihilation cross-section that are relevant for generic light WIMP candidates (see for instance~\cite{Papucci:2009gd,Abazajian:2010sq,Hutsi:2010ai,Abdo:2010dk,Mack:2008wu,Profumo:2009uf}). From the perspective of candidates suitable for CoGeNT-DAMA, this has been particularly emphasized in various recent works. To our knowledge, possible constraints from the isotropic diffuse gamma-ray emission or from the centre of our galaxy, are mentioned in~\cite{Fitzpatrick:2010em} and, respectively, in~\cite{Fitzpatrick:2010em,Feng:2008dz,Andreas:2008xy}. Also,  constraints from Milky Way Dwarf Spheroidal galaxies (dSph) are discussed in ~\cite{Fitzpatrick:2010em,Andreas:2010dz}.  In~\cite{Andreas:2010dz} in particular,  the analysis of the Fermi-LAT collaboration of the gamma-rays flux from Milky Way dSphs~\cite{Abdo:2010ex} has been exploited to set a conservative constraint on a 10 GeV scalar singlet candidate, and to tentatively exclude lighter candidates, assuming a Navarro-Frenk-White (NFW)~\cite{Navarro:1996gj, Navarro:2008kc} profile for the dark matter in dSphs. Constraints from antimatter in cosmic rays, in particular anti-protons and anti-deuterons, have been discussed in~\cite{Donato:2003xg,Bottino:2005xy,Nezri:2009jd,Bottino:2009km} and more recently in~\cite{Lavalle:2010yw}, where it is emphasized that candidates below $10$ GeV are disfavoured by observations of the anti-proton cosmic ray flux. Another potentially important signature is synchrotron radiation from the electrons and positrons produced directly, or indirectly through inverse Compton scattering, in the annihilation of dark matter (see for instance~\cite{Regis:2008ij,Hooper:2008zg,Borriello:2008gy,Crocker:2010gy}, and more in relation with CoGeNT-DAMA~\cite{Boehm:2002yz}). 

In the present article, we study further the constraints that may be set on light WIMP candidates related to CoGeNT-DAMA, based on the first-year Fermi-LAT  data on the spectrum of isotropic diffuse gamma-ray emission~\cite{Abdo:2010nz}. Our motivation for doing so is manifold.  The most obvious one is that we may anticipate that these observations are relevant for a generic light WIMP. In the analysis of~\cite{Abazajian:2010sq}, the first-year Fermi-LAT data are used to set 
constraints on the annihilation cross-section of generic candidates, and this down to $M_{DM} = 5$ GeV. For the lightest candidates annihilating into $\tau^+\tau^-$, the constraints touch the region relevant for a WIMP, $\svann \sim 3 \cdot 10^{-26}$~cm$^3$s$^{-1}$ (see Figure 2 of Ref.~\cite{Abazajian:2010sq}). In the analysis by the Fermi-LAT collaboration~\cite{Abdo:2010nz}, which covers  generic candidates with  10 GeV $\leq M_{DM} \leq 1$ TeV, the constraints appear to be more stringent and, depending on the astrophysical assumptions, some candidates may be excluded at 95 \% confidence level (C.L.). These works also have a focus on more ``popular'' DM candidates, like the neutralino,  Kaluza-Klein DM candidates or models designed to explain the possible excess of positrons in cosmic rays. 

The two analysis mentioned above essentially concur, and the differences in the strength of the constraints may be traced (albeit in a very convoluted way) to the use of distinct assumptions regarding the astrophysical uncertainties, in particular the distribution of inhomogeneities as a function of redshift. 
However, given the relevance of light WIMPs candidates for the anomalies reported by some direct detection experiments, and the current astrophysical uncertainties, we found of interest to make available yet another independent analysis, 
but  with a focus on the low mass region, 5 GeV $ \lsim M_{DM} \lsim 10$ GeV. Moreover the data from Fermi-LAT have the  potential to constrain candidates with a mass below 5 GeV, possibly down to $M_{DM} \sim$ 1 GeV. On one hand, going to lower masses is a tiny step, and, as a first approximation, one could simply extrapolate the existing constraints (from either~\cite{Abazajian:2010sq} or~\cite{Abdo:2010dk}). On the other hand, the current direct detection experiments are not sensitive to such low mass candidates (be them unrelated to CoGeNT-DAMA) and, in our opinion, it is of interest to be as precise as possible. A further motivation of ours is that we would like to confront two specific, albeit quite generic, models of dark matter, in particular a real scalar singlet, interacting through the Higgs portal, and a Dirac fermion, interacting through a new $Z^\prime$ ({\em i.e.} the $U(1)_Y$ portal). Although these models have similar pros (they are simple, minimal extensions of the Standard Model, hereafter SM, and they are directly relevant for CoGeNT-DAMA), and cons (SI elastic scattering with nuclei --- which is challenged by Xenon10/100--- and {\em ad hoc} mass scales and couplings), they have a distinct phenomenology with regard to  isotropic diffuse gamma-ray emission. Specifically, these candidates have distinct couplings to SM light degrees of freedom ({\em ie} $e^+/e^-$ and neutrinos) and thus have a distinct kinetic decoupling temperature, a parameter which is relevant to establish the mass of the smallest gravitationally bound DM objects that may form in the Universe, and consequently for the calculation of the diffuse isotropic gamma-ray flux. This implies that the scalar singlet (and its siblings with similar interactions) is sensibly more constrained than a candidate interacting through a $Z^\prime$, a feature which should be of interest for further model building. 
\bigskip

The plan of our article is consequently as follows. In section~\ref{sec:IGRB}, we briefly expose our hypothesis regarding the calculation of the isotropic diffuse gamma-ray emission. As in~\cite{Abdo:2010dk}, we will suppose that the contribution from dark matter annihilation within the halo of our galaxy (and possible subhalos within) is negligible, and so, that the dominant contribution is due to the isotropic gamma-ray background radiation (IGRB). For the sake of comparison with other approaches, and to set our conventions regarding the astrophysical parameters, we discuss in section~\ref{sec:RESULTS} the exclusion limits in general terms.  Last but not least, in section~\ref{sec:SCALAR}, we compare the outcome of our calculations for two minimal, toy models of dark matter which have been shown to be consistent with both CoGeNT-DAMA and WMAP, but have distinctive characteristics: a scalar candidate interacting through the Higgs portal, thus with `''Higgs-like'' couplings to ordinary matter, and a Dirac fermion interacting through a $Z^\prime$, hence with ``Z-like'' couplings. Our main results are given in Figures~\ref{fig:singlet} and~\ref{fig:singletSI} for the scalar singlet, and Figure~\ref{fig:psiSI} for the Dirac fermion. Finally we give our conclusion and prospects in section~\ref{sec:conclusions}.

\section{The isotropic extragalactic gamma-ray flux}
\label{sec:IGRB}

In this section we summarize the formalism we have used to compute the extragalactic contribution to the gamma-ray from dark matter annihilation. We follow the approach of~\cite{Cirelli:2009bb} and compare the result to other approaches in the next section. 

To begin with, we will work with the hypothesis that the contribution of dark matter annihilation within the halo of our galaxy to the diffuse isotropic gamma-ray emission is small compared with the extragalactic component, or IGRB, following the approach of~\cite{Abdo:2010dk}, to which we refer for future discussions on this point. 
 
The flux of gamma-ray today (in units of GeV$^{-1}$cm$^{-2}$s$^{-1}$sr$^{-1}$) from DM annihilation at any redshift $z$ is then given by~\cite{Ullio:2002pj}\footnote{The factor of $1/2$ is for a self-conjugate DM candidate. In the case of a Dirac fermion which we will also consider, there is an extra $1/2$ factor~\cite{Ullio:2002pj}.}
\begin{equation}
\label{eq:flux}
\frac{d\Phi_{\gamma}}{dE} = \frac{c}{4 \pi} \frac{\svann}{2 M^2_{DM}} \int_{0}^{\infty} dz^\prime \frac{1}{H(z^\prime) (1+z^\prime)^4} \frac{dN_{\gamma}}{dE'} \mathcal{B}^2(z^\prime){\rm e}^{-\tau(E',0,z^\prime) } \, .
\end{equation}
In this equation $H(z)(1+z)=H_0 h(z)(1+z)$ sets the relation between redshift interval and the proper distance interval, with $h(z) = \sqrt{\Omega_M (1+z)^3 + \Omega_{\Lambda}} $. For the cosmological parameters today we use the central value from the WMAP-7yrs measurements~\cite{Komatsu:2010fb}: $H_0= 70.4\  {\rm km/s/Mpc}$, $\Omega_M= 0.227$, $\Omega_b= 0.0456$ and $\Omega_{\Lambda}= 0.728$. A factor of $(1+z)^{-3}$ accounts for the dilution of the number density of photons from  expansion. Finally $E' = E (1+z^\prime)$ is the energy of the photon at redshift $z^\prime$ for an energy $E$ observed today. The factor $dN_{\gamma}/dE$ is the spectrum of gamma-rays produced in the annihilation of a pair of DM particle, at a rate given by $\svann$, and depends on the particle physics candidate. The function $\mathcal{B}(z)$ takes into account the density profile of dark matter at a given redshift $z$ and will be presented in Section~\ref{sec:boost}.
Finally the parameter $\tau(E^\prime,z,z')$ is the optical depth (of a gamma-ray emitted at $z^\prime$ at energy $E^\prime$ and observed at $z$), which takes into account the absorption that may occur on the path of a gamma-ray. For the case of a light WIMP ($M_{DM} \lsim 10-15$ GeV), and focusing on the Fermi-LAT window, which has measured the isotropic flux between $200$ MeV and $100$ GeV, the Universe is essentially transparent. This is an interesting simplification as it removes one source of uncertainty on the calculated gamma-ray flux, so we discuss this aspect in some details in the following section. 

\begin{figure}[b!]
\centering
\includegraphics[width=0.5\columnwidth]{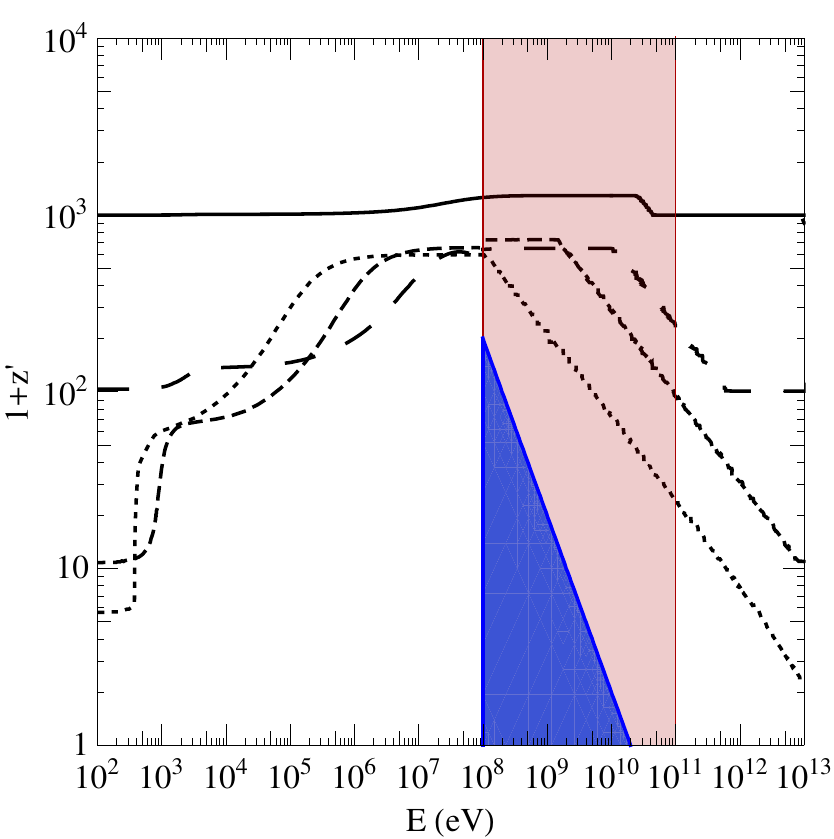}
\caption{Contour plot of the optical depth with $\tau(E',z,z')=1$, where $E'/(1+z')=E/(1+z)$. Here $E$ is the energy observed at redshift $z$, so that $E^\prime$ is the emission energy at redshift $z^\prime$. The different lines correspond to the different observers. For each, they give the maximal possible redshift at emission: $z= 0$ dotted line, $z=10$ dashed curve, $z=100$ long dashed line and $z=1000$ solid line. The red (gray) region denotes the Fermi-LAT measured energy range. The triangular region gives the relevant (for observations today)  redshifts  for a DM candidate of mass $M_{DM} = 20$ GeV.}
\label{fig:op}
\end{figure}

\subsection{Optical depth}\label{sec:op}
For increasing energy, the  dominant processes that may affect the propagation of a gamma-ray between redshift $z$ and today are (see, for instance,~\cite{Slatyer:2009yq})
\begin{eqnarray}
{\rm Photo-ionization} & \qquad \gamma + H (He, He^{+}) \rightarrow e^{-} + H^{+} (He^{+}, He^{++})\,, \label{pi}\\
{\rm Compton\  Scattering}  & \qquad   \gamma + e^{-}\rightarrow  \gamma + e^{-}\,,\label{cs}\\
e^+e^-{\rm pair \  production\  on\  matter} & \qquad \gamma + A \rightarrow A+ e^+ + e^-\,, \label{ppm}\\
{\rm Photon-photon\  scattering}  &  \qquad  \gamma +\gamma_{{\rm BkGrd}} \rightarrow \gamma + \gamma  \,, \label{ps}\\
\qquad e^+e^-{\rm pair\  production}  & \qquad \gamma + \gamma_{{\rm BkGrd}}  \rightarrow e^+ + e^- \label{ppcmb}\,.
\end{eqnarray}
To compute  the processes (\ref{pi}) and (\ref{ppm}) we have used the approach of~\cite{Slatyer:2009yq}, for Compton scattering we refer to~\cite{Blumenthal:1970gc}, to~\cite{Agaronyan} for  pair production on a background photon and, finally to~\cite{Zdziarski:1988pb} for the photon-photon scattering contribution. 

The results of our calculations are summarized in Figure~\ref{fig:op}, and they are in agreement with, for instance,~\cite{Hutsi:2010ai}. There we show, in function of the observed energy today (dotted at $z=0$), the redshift $1+z^\prime$ at which the optical depth was $\tau(1+z^\prime) =1$, such that the Universe may be considered to be optically thin (transparent) at lower redshifts. 
The red (gray) region corresponds to the energy window of Fermi-LAT, which shows that only the (\ref{ppm} - \ref{ppcmb})  processes are relevant (see also~\cite{Hutsi:2010ai}). The triangular region (in blue/dark gray in the plot) within the Fermi-LAT window gives the allowed  redshifts at emission in function of energy for a candidate with mass  $M_{DM} \lsim 20$ GeV. The fact that $\tau(1+z') \ll 1$ within the triangular means that the Universe is optically thin for these gamma-rays. However, we show in the plot only the effect from scattering or pair creation on the photons from the cosmoc microwave background (CMB), as is relevant for large redshifts, but, at lower redshifts, $z \lsim 6-10$, we should also take into account the ultraviolet (UV) photons produced by the formation of the first stars (an effect not shown in Figure~\ref{fig:op}). Unlike the CMB, there is much uncertainty regarding the distribution and spectrum of this extra background light. Here we follow~\cite{Hutsi:2010ai} and also~\cite{Gilmore:2009zb} and~\cite{Stecker:2005qs}, where  the possible impact of the low $z$ UV background is discussed. In~\cite{Hutsi:2010ai}, three models for the UV background are considered. A glance at Figure 3 in Ref.~\cite{Hutsi:2010ai} reveals that, for the case of a so-called realistic photon background, the effect is not relevant for a light WIMP.  A similar conclusion may be drawn from the plots in Figure 11 of~\cite{Gilmore:2009zb}, which shown that absorption due to UV photons is negligible within the range relevant for a light WIMP within the Fermi-LAT window. Absorption due to UV photons is more important according to the somewhat older model of~\cite{Stecker:2005qs}, but, keeping in mind the fact that spectrum of photon produced by a, say, $20$ GeV candidate typically peak at $1$ GeV, we may safely conclude from this reference that absorption is essentially negligible, even in the most pessimistic scenario, if we concentrate on a light WIMP and the Fermi-LAT data. Thus, in the calculation of the gamma-ray flux, the only relevant astrophysical uncertainty  is the distribution of dark matter halos as a function of redshift, which we now discuss.

\subsection{Enhancement factor from structure formation}\label{sec:boost}

To compute the extragalactic contribution to the gamma-ray flux from dark matter annihilation, we need the distribution of dark matter halos as a function of redshift. There are different approaches to this complicated problem. Needless to say, this will be the main source of uncertainty on our results. Here we found convenient to follow a standard, semi-analytical approach based on the Press-Schechter formula. In particular we follow the notation of~\cite{Cirelli:2009bb}, which is itself based on~\cite{Natarajan:2008pk}. One should bare in mind that different approaches may give substantially different results.  For a review on halo models and structure formation we refer to~\cite{Cooray:2002dia,Barkana:2003qk}.

As the Universe expands, the initially small perturbations in the dark matter distribution evolve in a non-linear way into a complex network of structures made of dark matter halos of various scales. In the $\Lambda$CDM model for structure formation, dark matter halos are assumed to form hierarchically bottom-up, with small structures merging into larger halos, in a self-similar pattern. Under these hypotheses, the Press-Schechter~\cite{Press:1973iz} empirical formula gives an  analytical expression for the abundance of dark matter halos of mass $M$ at a given redshift $z$, on the basis of the predictions of linear perturbation theory  for the power spectrum of perturbations:
\begin{equation}
\label{eq:ps}
\frac{dn}{dM}(z,M) = \sqrt{\frac{2}{\pi}} \frac{\rho_{c}\Omega_{DM}}{M}  \frac{d \sigma(R,z)}{dM} \frac{\delta_c}{\sigma^2(R,z)} {\rm exp}\Big(-\frac{\delta_c^2 }{2 \sigma^2(R,z)  }\Big)\,,
\end{equation}
with $\rho_c$ being the critical density. The factor $\delta_c$ Eq.~(\ref{eq:ps}) is the critical over-density required in a spherical model for gravitational collapse. We take $\delta_c=1.28$ following~\cite{Barkana:2003qk}. The other key quantity is $\sigma^2(R,0)$,
which is the variance of the density field, obtained using linear perturbation theory, in a sphere of radius $R$ containing a mean mass $M$. Its evolution  with respect to redshift is simply given by the growth factor, using the standard factorization between scale dependence and redshift~\cite{Eisenstein:1997jh}. The variance of the density field on a scale $R$ is related to the matter power spectrum $P(k)$ through convolution with a spherical top-hat window function $W(kR)$, with $R^3= 3 M/4 \pi \rho_c \Omega_M$. To compute  the linear matter power spectrum today, we have used  CAMB~\cite{Lewis:1999bs,CAMB_code}, linearly extrapolated at very large wave-number $k$ for the evaluation of Eq.~(\ref{eq:ps}). Both $P(k)$ and $\sigma$ are normalized by computing $\sigma$ in a sphere of $R= 8 h^{-1} \rm Mpc$, with
$\sigma_8=0.8$~\cite{Natarajan:2008pk,Ullio:2002pj}. 

From the distribution of halo of mass $M$ at redshift $z$, we may compute the flux using 
\begin{eqnarray}\label{eq:b1}
\mathcal{B}^2(z) & = & \mathcal{\bar{B}}^2(z) + B^2(z)\,, \nonumber\\
& = & \rho^2_c \Omega_{DM}^2 (1+z)^6 \Big( 1+ \frac{1}{\rho^2_c \Omega_{DM}^2 (1+z)^3}\int dM \frac{dn}{dM}(z,M) (1+z)^3 \int_0^{r_{vir}(M,z)}
  dr \ 4 \pi r^2 \rho^2(r,M) \Big)\,,
\end{eqnarray}
where the first term is the contribution from the smooth distribution of dark matter, which in practice is negligible, and $\rho(r,M)$ is the dark matter density profile of dark matter in halo. The integral over $\rho$ is calculated with a cut-off given by the virial radius $r_{vir}(M,z)$, which is usually expressed in terms of the  so-called concentration parameter $c_{vir}= r_{vir}(M,z)/r_s$ where $r_s$ is the core radius which depends on the choice for the profile~\cite{Bullock:1999he}.  The virial radius of a halo of mass $M$ at a given redshift $z$ is itself defined as the radius within which the mean density of the halo is $\Delta_{vir}$ times the smooth density $\rho_c\Omega_M$. In the $\Lambda$CDM model, this virial over-density is given by
\begin{equation}
\Delta_{vir}(z) = \frac{18 \pi^2+82 (\Omega_M(z)-1)-39 (\Omega_M (z)-1)^2}{\Omega_M(z)}\,.
\end{equation}
which generalizes the standard result $\Delta_{vir}(z) = 18 \pi^2 \sim 200$ of a CDM Universe. 

As is standard practice for primordial halos, we consider throughout the paper the NFW profile as a benchmark~\cite{Navarro:1996gj, Navarro:2008kc}.
The NFW density profile has a simple analytic expression in terms of the concentration parameter
\begin{equation}
F_{\mbox{\rm \tiny NFW}}(c_{vir})= \frac{c_{vir}^3}{3} \Big( 1-\frac{1}{(1+c_{cvir})^3}\Big) \Big( {\rm Log}(1+c_{vir})-\frac{c_{vir}}{(1+c_{vir})}\Big)^{-2}\,. 
\end{equation}
Using the previous definitions, the integral over the halo mass function can finally be recast into the form
\begin{equation}
\label{eq:boost}
B(z) = \frac{\Delta_{vir}}{3 \rho_c \Omega_M} \int_{M_{\rm min}}^{\infty} dM  M \frac{dn}{dM} F_{\mbox{\rm \tiny NFW}}(c_{vir}(z,M))\,,
\end{equation}
where $M_{\rm min}$ is the minimum halo mass (see Section~\ref{sec:RESULTS}) so that Eq.~(\ref{eq:b1}) becomes:
\begin{equation}
\label{eq:total}
\mathcal{B}(z) =  \rho_{c}\Omega_{DM} (1+z)^3 \sqrt{1+{\rm B}(z)}\,.
\end{equation}
This is the enhancement factor which appears in the extra-galactic flux, Eq.~(\ref{eq:flux}). The dark matter distribution is described by the mean cosmological matter density $\mathcal{
\bar{B}}$ plus the sum over all average enhancements due to a halo of mass $M$, weighted over the mass function. The abundance and distribution of halos depend primarily on halo mass and the concentration parameter itself depends on the halo mass at a given redshift. Therefore, the range of $M_{\rm min}$ and the functional form of $c_{vir}$ are the major sources of uncertainty in the determination of the diffuse gamma-ray flux and we will discuss our choices in Section~\ref{sec:RESULTS}.

\subsection{Gamma-ray spectrum at production}
\begin{figure}[t!]
\centering
\includegraphics[width=0.5\columnwidth]{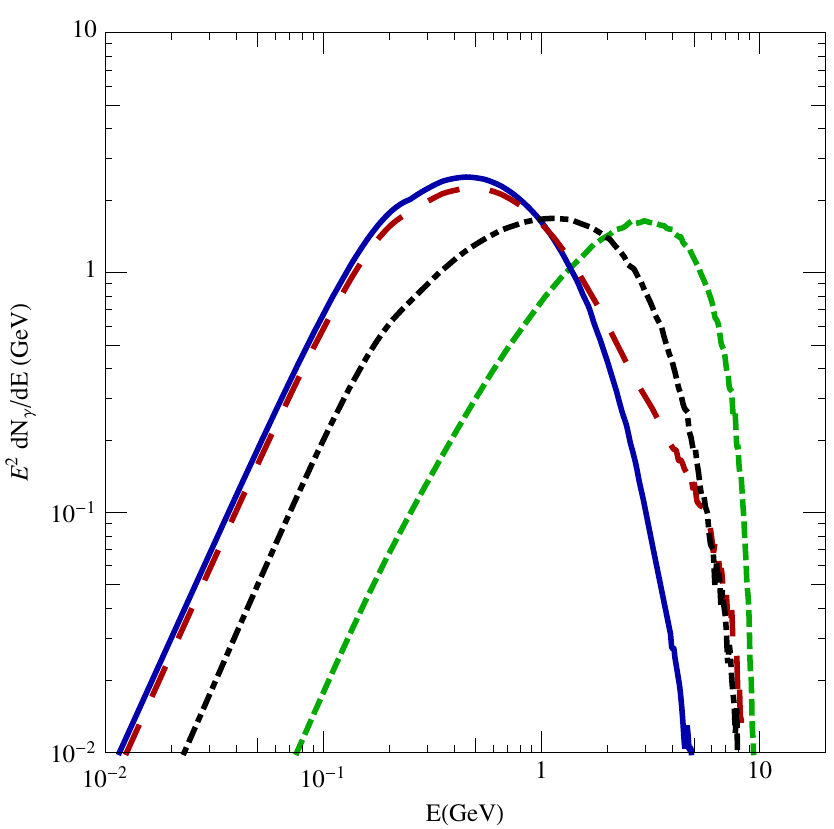}
\caption{Example of $E^2 dN_{\gamma}/dE$ gamma-ray spectra from the annihilation of a 10 GeV dark matter particle, as a function of the photon energy $E$. The solid blue line represents a pure annihilation into $b\bar{b}$ ($BR_{b\bar{b}}=100\%$), the green dashed curve is for a pure annihilation into $\tau^+\tau^-$ ($BR_{\tau^+\tau^-}=100\%$), the red long dashed to a ``Higgs-like" annihilation ($BR_{\tau^+\tau^-}\sim 9\%$ and $BR_{b\bar{b}} \sim 83\%$), while the black dot-dashed line refers to a``Z-like" annihilation ($BR_{l^\pm}\sim 3\%$ and $BR_{q\bar{q}}\sim 13\%$).}
\label{fig:spectra}
\end{figure}

The last  ingredient we will need to compute the contribution to the IGRB, is the spectrum of gamma-rays produced per DM annihilation,
\begin{equation}
\frac{dN_{\gamma}}{dE} = \sum_f \frac{dN_{\gamma}^f}{dE} BR_f\,,
\end{equation}
where $dN_{\gamma}^f/dE$ is the photon spectrum for the $f$ annihilation final state with branching ratio $BR_f$. The gamma-ray spectra from hadronization and final state radiation are calculated with Pythia 8.1~\cite{Sjostrand:2007gs}. In Figure~\ref{fig:spectra} we show a few spectra representative of the particle models that we will analyze. The solid blue line is the final state radiation from annihilation into $b\bar{b}$ pairs, the green short-dashed curve comes from a $BR=100\%$ into $\tau^+\tau^-$. As is well known, in both cases, the main contribution comes from the decay of neutral pions into gammas, produced from the hadronization of the $b\bar{b}$ pair or from the semi-hadronic decay of the $\tau^+\tau^-$. The most relevant feature for our analysis, which will make  use of the lower energy edge of the Fermi-LAT data, is that the photon spectrum from $\tau^+\tau^-$  is harder than the one from $b\bar b$ quarks. The red long-dashed curve is the photon spectrum for light WIMP with ``Higgs-like" coupling to SM degrees of freedom, namely with a branching ratio dictated by the Yukawa couplings. For instance, for a scalar dark matter particle with $M_{DM}= 10$ GeV, it corresponds to a $BR_{b\bar b}=83\%$ and to a $BR_{\tau^+\tau^-}= 9\%$. The black dot-dashed line is the spectrum for a DM particle with ``Z-like'' couplings, in which case $BR_{b\bar b} \sim 3\times BR_{l^\pm} \sim 13\%$.

\section{General results regarding a light WIMP}
\label{sec:RESULTS}
In this section we discuss, and compare to known results, our choice for astrophysical quantities like the concentration parameters and the minimum DM halo mass that may form. These parameters are defined in a non-unique way in the literature, and the choice of their parameterization will affect the limits on the annihilation cross-sections. Once the technical details are worked out, we present some generic constraints relevant for a light WIMP, with $M_{DM}$ up to 20 GeV.

\subsection{Uncertainties from structure formation}
\label{subsec:hmuncert}
\begin{figure}
\begin{minipage}[t]{0.5\textwidth}
\centering
\includegraphics[width=0.9\columnwidth]{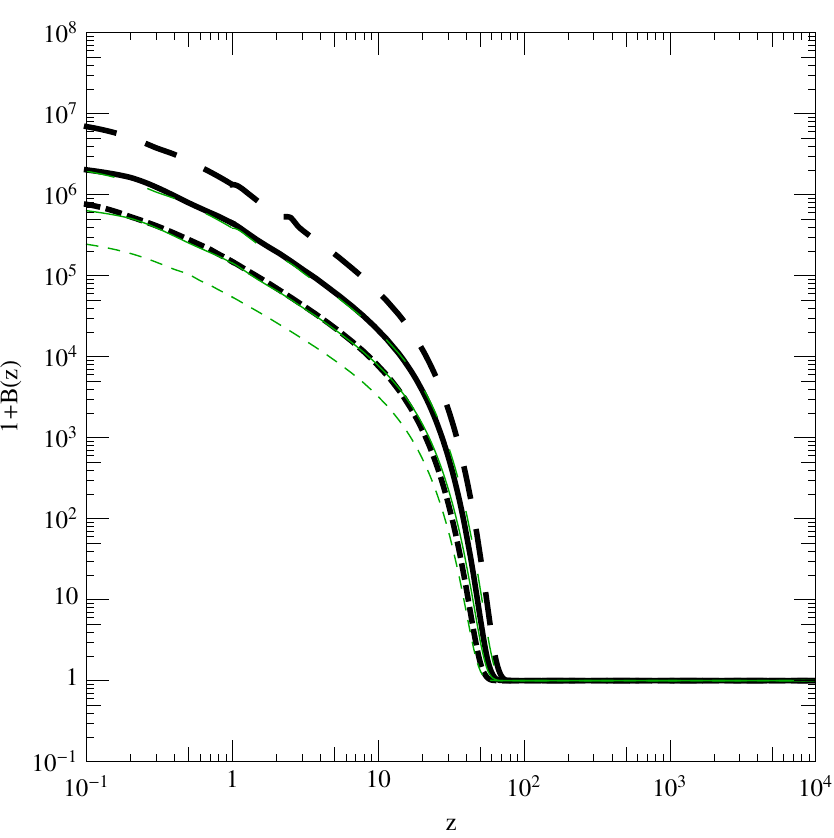}
\end{minipage}
\hspace*{-0.2cm}
\begin{minipage}[t]{0.5\textwidth}
\includegraphics[width=0.9\columnwidth]{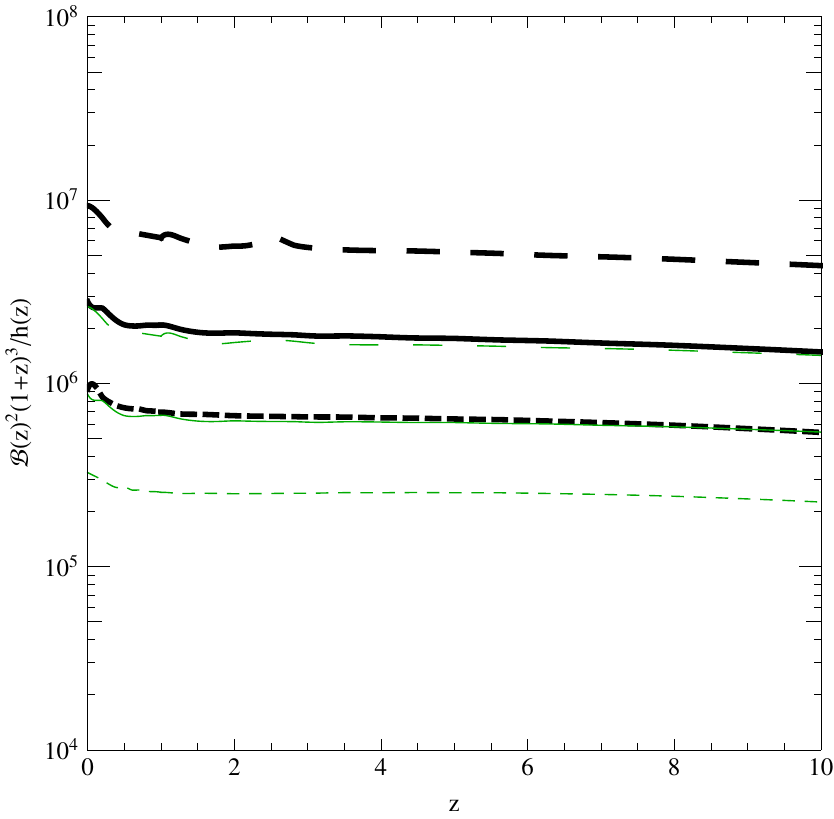}
\end{minipage}
\caption{On the left, boost factor $1+{\rm B}(z)$ as a function of the redshift $z$. On the right, zoom at small redshift values of the total enhancement factor $\mathcal{B}^2(z)/h(z)(1+z)^3$. A NFW density profile for the dark matter is assumed. The thin green lines are given for the $C_{\rm WMAP}$ concentration parameter~\cite{Maccio:2008xb}, while the thick black lines are for the $C_{\rm PL}$ concentration parameter~\cite{Bullock:1999he}. The long dashed lines refer to $M_{\rm min}= 10^{-8} M_{\odot}$, the solid to $M_{\rm min}= 10^{-6} M_{\odot}$ and the short dashed to $M_{\rm min}= 10^{-4} M_{\odot}$.}
\label{fig:boost}
\end{figure}

Simulation of halos~\cite{Navarro:1996gj,Bullock:1999he} indicate that there is a strong correlation between the concentration parameter and the halo mass $M$. Among other properties, they are inversely related, {\it e.g.} higher mass halos are less concentrated. A model that fits the results of the simulations is:
\begin{equation}
c_{vir}(z,M) = \frac{c_{vir} (0,M)}{1+z}\,,
\end{equation}
meaning that $c_{vir}$ is inversely proportional to the redshift as the radius of a halo increases as the Universe expands. We use two different power law fits for $c_{vir}$. The first one ($C_{\rm WMAP}$ hereafter) is the WMAP-5yrs best fit in~\cite{Maccio:2008xb}, which reads ${\rm Log}\ c_{vir} = 0.971 - 0.094\  {\rm Log}\  [M_{vir}/(10^{12} h^{-1} M_{\odot})]$. The second one ($C_{\rm PL}$ from now on) follows a simple power-law behavior, $c_{vir} \propto M^{-0.1}$, as indicated by numerical simulations~\cite{Bullock:1999he} and is normalized at the same mass scale as $C_{\rm WMAP}$.

\bigskip 
The other quantity that affects the final results, namely the upper bounds on $\svann$, is the choice of the minimal halo mass, appearing as the lower limit in the integral of Eq.~(\ref{eq:boost}). It has been shown in~\cite{Hofmann:2001bi,Schwarz:2001xj} that free streaming and  collisional damping of a WIMP with particles from the thermal bath lead to a small-scale cut-off in the DM density perturbation power spectrum.  The size of the smallest gravitationally bound structure is set by the properties of the WIMP. After the DM particle abundance has frozen out in the early Universe, it may still stay in local thermal equilibrium by elastic scattering processes with SM particles, $DM + f \rightarrow DM + f$. As the Universe expands the elastic scattering rate decreases and the WIMPs kinetically decouple at some temperature, $T_{\rm kd}$ which depends on the WIMP properties. From this moment on the WIMPs can stream freely and wash out matter contrasts on small scales. For a candidate like a neutralino, the complete decoupling from the thermal bath typically happens when the temperature has dropped by a factor $\sim 10-1000$ with respect to the chemical decoupling temperature, namely $T_{\rm kd}\sim\mathcal{O}({\rm MeV})$ \cite{Hofmann:2001bi,Schwarz:2001xj}. 
In the next section we will consider two distinct light WIMP candidates, a Dirac fermion with ``Z-like'' couplings to SM degrees of freedom, and a scalar candidate with ``Higgs-like'' couplings to ordinary matter, {\em i.e.} through Yukawa couplings. Here we estimate the temperature of kinetic coupling and the resulting estimates for the minimum halo mass, emphasizing the differences between the two models. 

In the case of a scalar DM interacting through the Standard Model Higgs, we could expect that kinetic decoupling occured at about the same temperature as chemical decoupling, typically $T_{\rm kd}\gsim$ 100 MeV for the candidates of interest here. More specifically, for SM Yukawa couplings, the most relevant  scattering is {\em a priori} with $c$, $b$ and $\tau^-$ and their antiparticles, which are non-relativistic, and thus have an exponentially suppressed abundance at the time of interest. As a first estimate for $T_{\rm kd}$, we may balance the rate for elastic scattering with the expansion rate of the Universe. In this approximation  kinetic decoupling is  like chemical decoupling, but the density of dark matter is replaced by the density of target fermions in the collision rate. From this analogy, we estimate that kinetic decoupling occurs for $x_{\rm kd} = m_f/T_{\rm kd} = {\cal O}(20)$ assuming  $\sigma= {\cal O}(10^{-36}$ cm$^2)$, which gives $T_{\rm kd} \sim 250$ MeV in the case of $f=b/\bar b$, while for $c/\bar c$ and $\tau^+/\tau^-$ we have $T_{\rm kd} \sim 100$ MeV. These temperatures are of the order of the chemical decoupling ($fo$) temperature, $T_{fo} \sim 250-500$ MeV for $M_{DM} \sim 5-10$ GeV. Things are however a bit more subtle, as momentum exchange typically requires that many collisions take place, $N_{\rm col}\gsim 1$. From the analysis of Ref.~\cite{Bringmann:2009vf}, we have  $N_{\rm col} \sim M_{DM}/T \gg 1$ or $N_{\rm col} \sim M_{DM}/m_{f}\gsim 1$ for relativistic or non-relativistic targets respectively. Furthermore the interaction with $\mu^+/\mu^-$ are small but still substantial, so we need a more precise formulation. For this, we rely on the results of  Ref.~\cite{Bringmann:2009vf} where a Boltzmann equation for the kinetic decoupling temperature $T_{\rm kd}$ has been derived for general targets (but provided $M_{DM} \gsim m_f$).\footnote{The Boltzmann equation for the kinetic decoupling temperature derived in~\cite{Bringmann:2009vf} is valid for both relativistic and non-relativistic targets, but rests on an expansion of the collision term to leading order in parameters like $\omega/M_{DM}$. One should  keep in mind that this approximation may not be very good for the heaviest targets we consider, in particular $b\bar b$ quarks but, for the case of ``Higgs-like'' couplings,  we only need to assume that it applies for collisions with lighter, albeit non-relativistic targets, which seems reasonable.} From the Boltzmann equation given by Eq.(10) (and the results in Appendix A) of~\cite{Bringmann:2009vf} we infer that departure from kinetic equilibrium between the scalar DM candidate and a specific fermionic species  occured when a kinetic collision term, $c(T)$,  was of the order of the expansion rate $H(T)$,
\begin{equation}
M_{DM} c(T_{\rm kd}) \sim {\tilde g_{\rm eff}^{1/2} \over (2 \pi)^3 M_{DM}^3 T_{\rm kd}}\, {\lambda^2 m_f^2\over m_h^4} \int dk k^5 \omega(k) e^{-\omega(k)/T_{\rm kd}} \sim H(T_{\rm kd}).
\label{eq:Htkd}
\end{equation} 
Here $\lambda$ is the scalar-Higgs coupling, $m_h$ the Higgs mass, $\omega(k) = \sqrt{k^2 + m_f^2}$ and $\tilde g_{\rm eff}^{1/2} = g_{\rm eff}^{1/2}(T)/(1+T\,g_{\rm eff}^\prime/4 g_{\rm eff})$ where $g_{\rm eff}$ is the number of degrees of freedom contributing to the energy density at kinetic decoupling.\footnote{The parameter $\tilde g_{\rm eff}$ in (\ref{eq:Htkd}) may differ from $g_{\rm eff}$ by a factor of ${\cal O}(2)$ between $T=100$ and 200 MeV (see Figure 1 in Ref.\cite{Bringmann:2009vf}), an effect we have however neglected in our order-of-magnitude estimates for $T_{\rm kd}$.} We find that kinetic decoupling occured at $T_{\rm kd} \sim 300$ MeV in the case of a 10 GeV candidate scattering with $b/\bar b$ quarks, while we find $T_{\rm kd} \sim 100-170$ MeV for $c/\bar c$ quarks and $\tau^+/\tau^-$  respectively in the case of $M_{DM}=$ 5 and 10 GeV candidates, values which are not too far from our naive estimates. As for the muons, their coupling to the Higgs is parametrically smaller, but there are also lighter. In this case we find  $T_{\rm kd} \sim 50-150$ MeV, again for 5 and 10 GeV candidates. For lighter SM species, the decoupling occurs at much larger temperatures (we get for instance $T_{\rm kd} \sim 700$ MeV for $e^+/e^-$). Thus we take  $T_{\rm kd} \sim 50-150$ MeV as our estimate of the range of kinetic decoupling temperatures of the scalar DM candidate interacting through the Standard Model Higgs.\footnote{We notice that these estimates imply that our candidates decoupled around the QCD phase transition. Beside the fact that one may be uncomfortable with considering free quarks below $T \sim 1$ GeV (an issue we do not address here), the QCD phase transition may also have an interesting impact on the formation of the smallest dark matter halos, and issue that has been addressed in, for instance, Ref.~\cite{Schmid:1998mx}. A relevant effect of the QCD phase transition is that there is a sharp drop of the speed of sound in the baryon-photon fluid. For dark matter that has kinetically decoupled {\em before} the QCD phase transition, this may lead to the formation of dark matter halos as light as $10^{-10} \, M_\odot$ or even less~\cite{Schmid:1998mx}. Whether this effect may be relevant for the candidates with ``Higgs-like'' couplings is unclear, given the proximity of the QCD phase transition critical temperature $T_c \sim 170$ MeV and the kinetic decoupling temperature, so we leave this interesting possibility for a possible future study. }

 In the case of ``Z-like" we expect a priori $T_{\rm kd}$ to be $\sim $ 1-10 MeV  from previous estimates, like those based on the neutralino,  since in this case the collisions with $\nu/\bar\nu$ and $e^+/e^-$ from the thermal bath are most  important. To begin with we refer to the analysis of Ref.~\cite{Green:2005fa}, where an explicit formula for $T_{\rm kd}$ is given for the case of scattering of WIMPs on light, relativistic species in the thermal bath. Concretely, the thermal averaged elastic cross-section on relativistic light fermions is expanded as $<\sigma_{\rm el}> = \sigma_0^{\rm el}\  (T/M_{DM})^{1+l}$ and the dependence on the temperature is parameterized by $l$. In the framework of Ref.~\cite{Green:2005fa} kinetic decoupling is assumed to take place when the rate of collisions is of the order of the expansion rate, hence this approach is similar to the one we have advocated when deriving our naive estimates of the kinetic decoupling for the scalar interacting through the Higgs.  In the Dirac fermion case, the elastic cross-section between the DM fermion and the SM fermions is mediated by the $Z'$, is S-wave and  proportional to $\sigma_0^{\rm el} \sim \sin\phi^2 G^2_F M_{DM}^2 m^4_W/m^4_{Z'} $. The  factor $\sin\phi$ of suppression with respect to the elastic cross-section mediated by a $Z$ boson is proportional to the kinetic mixing parameter $\epsilon$ and to the mass of the $Z'$ boson, as defined in Ref.~\cite{Mambrini:2010dq}. Substituting $\sigma_0^{\rm el}$ in Eq.(4) of~\cite{Green:2005fa}, the kinetic decoupling temperature is given by
\begin{equation}
T_{\rm kd} \sim \big[\frac{\zeta(3)}{\pi^2}\big(\frac{90 g_{\rm eff}^{1/2}}{8 \pi^3}\big) M_{Pl} \frac{G^2_F m^4_W}{m^4_{Z^\prime}} \sin\phi^2\big]^{-1/3}\,.
\label{eq:tkdZp}
\end{equation}
Notice that $T_{\rm kd}$ does not depend on the DM mass but depends on the details of the $Z'$ model, specifically on the mass of the hidden gauge boson and the kinetic portal. In our estimation we use $m_{Z'} = 15$ GeV and $\epsilon$ in the range $5 \times 10^{-4} - 5 \times 10^{-3}$, values which have been shown in Ref.~\cite{Mambrini:2010dq} to give the observed relic abundance for the light DM candidate. In this case $T_{\rm kd}$ is 5 - 25 MeV, as reported in Table I, slightly colder than in the case of the singlet scalar, $T_{\rm kd} \sim 50-150$ MeV. For the sake of comparison, we have also tentatively estimated the kinetic decoupling temperature using the equivalent of Eq.~(\ref{eq:Htkd}) for a Dirac fermion interacting through a $Z^\prime$ and found reasonable agreement, with $T_{\rm kd} \sim 5-15$ MeV, hence we adopt here the range $T_{\rm kd} \sim 5-25$ MeV for our estimates of the minimum halo masses for the Dirac fermion DM.

\bigskip
From the kinetic decoupling temperature,  one may estimate the free streaming length $k_{\rm fs}$ and the mass of the lightest bound structure, or dark matter halo, $M_{\rm min} = 4/3\  \rho_{DM} (\pi/k_{\rm fs})^3 \,M_\odot$. We use here the results from Refs.~\cite{Green:2005fa,Bringmann:2009vf} and Ref.~\cite{Loeb:2005pm}, which give slightly different answers. It is beyond our scope to choose between these options, so we give estimates for both cases. From the analysis in Ref.~\cite{Bringmann:2009vf}, one has
\begin{equation}
\label{eq:minhalomass}
M_{\rm min} \approx 2.9 \times 10^{-6} \left(\frac{1+ \ln (g_{\mbox{\rm eff}}^{1/4} T_{\mbox{\rm kd}}/50\, \mbox{\rm MeV}) /19.1}{(M_{\mbox{DM}}/100\, \mbox{\rm GeV})^{1/2} g_{\mbox{\rm eff}}^{1/4} (T_{\mbox{\rm kd}}/50\, \mbox{\rm MeV})^{1/2}}\right)^3 M_\odot.
\end{equation}
Using this expression, we obtain $M_{\rm min} \sim 10^{-4} \,M_\odot$ for  a ``Z-like" candidate, and $M_{\rm min}\sim 10^{-5} \,M_\odot$  for ``Higgs-like'' couplings (see Table I). For an alternative approach, we consider the results of Ref.~\cite{Loeb:2005pm}, and in particular
\begin{equation}
M_{\rm min} \approx 10^{-4} \Big(\frac{T_{\rm kd}}{10\  \rm{MeV}}\Big)^{-3} M_{\odot}\,,
\label{eq:loeb}
\end{equation}
an expression which was obtained based on a numerical evaluation of $k_{\rm fs}$. Using our estimates for the kinetic decoupling temperature, we obtain halo masses that are typically lighter for ``Higgs-like'' couplings, possibly as small as $M_{\rm min} \sim 10^{-8} \,M_\odot$, than for ``Z-like'' couplings, with $M_{\rm min} \sim 10^{-6} \,M_\odot$ respectively. We resume in Table I typical possible values of $T_{\rm kd}$ and $M_{\rm min}$  for a $M_{DM}= 5$ GeV and 10 GeV DM candidates, both with WMAP abundance. To emphasize the possibility that the scalar may potentially form lighter halos, in the evaluation of the diffuse gamma-ray signals we consider three different minimum mass values: $M_{\rm min}=10^{-8},10^{-6} \,M_\odot$ and $10^{-4} \,M_\odot$ for the model independent and for the singlet case, while only two values for the ``Z-like" DM candidate, $M_{\rm min}=10^{-6} \,M_\odot$ and $10^{-4} \,M_\odot$. For these models the minimum mass allowed by acoustic oscillations is far below the smallest halo mass permitted by the free streaming length scale $k_{\rm fs}$~\cite{Bringmann:2009vf}. Also, it is perhaps worth keeping in mind that some DM models designed to explain CoGeNT/DAMA may require different couplings than that of the Standard Model Higgs or the simple $Z^\prime$ model discussed here, opening the possibility of an even higher kinetic decoupling temperature, and thus lighter dark matter halos.

\begin{table}[t]
  \begin{center}
    \begin{tabular}{c|c|c|c}
Model & $T_{\rm kd}$ (MeV)  &\multicolumn{2}{|c}{ $M_{\rm min}$  ($M_{\odot}$) }   \\
      \hline\hline
      ``H-like" &  50-150 &  $5\cdot 10^{-5}-3\cdot 10^{-6}$ & $8\cdot 10^{-7}-3\cdot 10^{-8}$\\
      \hline
      ``Z-like" & 5 - 25  & $1 \cdot 10^{-3} - 5\cdot  10^{-5}$ & $8\cdot 10^{-4} - 6\cdot 10^{-6}$\\
            \hline\hline
    \end{tabular}    
       \label{tab:tkd}
  \end{center}
\caption{Range of kinetic decoupling temperatures and the corresponding minimal halo masses, $M_{\rm min}$, for ``Z-like" and ``Higgs-like'' DM candidates (see text). The extreme values for $M_{\rm min}$ correspond to candidates with mass of $5$ GeV (lower) and $10$ GeV (upper). Regarding $M_{\rm min}$, the results from Eqs.~(\ref{eq:minhalomass}) and~(\ref{eq:loeb}) are reported respectively in column 2 and 3.}
\end{table}

\bigskip

We finally show in Figure~\ref{fig:boost} the boost, Eq.~(\ref{eq:boost}), and total enhancement factor, Eq.~(\ref{eq:total}), as a function of the redshift $z$. The (green) thin lines are for the $C_{\rm WMAP}$ concentration parameter, while the (black) thick curves match the simple power law behavior for $c_{vir}$. The long-dashed lines correspond to $M_{\rm min}=10^{-8} \,M_\odot$, the solid curves to $M_{\rm min}=10^{-6} \,M_\odot$ and finally the short-dashed to $M_{\rm min}=10^{-4} \,M_\odot$ (the same convention for the curves description is maintained throughout the paper). On the left panel, we see that the enhancement due to structure formation starts to dominate over the smooth component at $z \leq 30$. The boost depends crucially on the concentration parameter, as $C_{\rm WMAP}$ leads to smaller boosts with respect to $C_{\rm PL}$; this has already been shown in~\cite{Huetsi:2009ex}, and we found agreement with their results. In the case of the $C_{\rm WMAP}$ parameterization we have checked that our results are equivalent to those in Fig.~(1) of Ref.~\cite{Cirelli:2009bb}.
On the right-hand side we show a zoom of the total dark matter density at small redshifts, the same quantity represented in Fig.(1) of Ref.~\cite{Abdo:2010dk}. The Fermi-LAT collaboration has used a different approach to compute the dark matter halo distribution, however we may compare their analytical estimation (labeled {\em BulSub}) with our calculations with $M_{\rm min}=10^{-6} \,M_\odot$: for $C_{\rm WMAP}$ we are 10$\%$ below their estimation, while for $C_{\rm PL}$ our boost is a factor $\sim 3$ above their curve. It is this small difference in the boost value at small redshifts (which are the values that contribute the most to the diffuse photon flux for a light WIMP) that will explain our more stringent upperbounds. We note that for a given concentration parameter and $M_{\rm min}=10^{-6} \,M_\odot$, the value $M_{\rm min}=10^{-8}(10^{-4}) \,M_\odot$ corresponds to increase (decrease) the boost by a factor $\sim 3$. Even though the diffuse photon flux is also increased or decreased by the same amount, below we show the upper bounds for all the $M_{\rm min}$ cases and for the two fit of the $c_{vir}$.

One should  keep in mind the uncertainty related in the choice of profile, here a NFW. Less cuspy profiles, like Einasto~\cite{Graham:2005xx} or Burkert~\cite{Burkert:1995yz}, will give a smaller boost, typically by a factor of 3 in the former case, and a factor of 8 in the latter~\cite{Cirelli:2009bb}. 


\subsection{Discussion on the upper bounds}
\begin{figure}
\begin{minipage}[t]{0.5\textwidth}
\centering
\includegraphics[width=0.9\columnwidth]{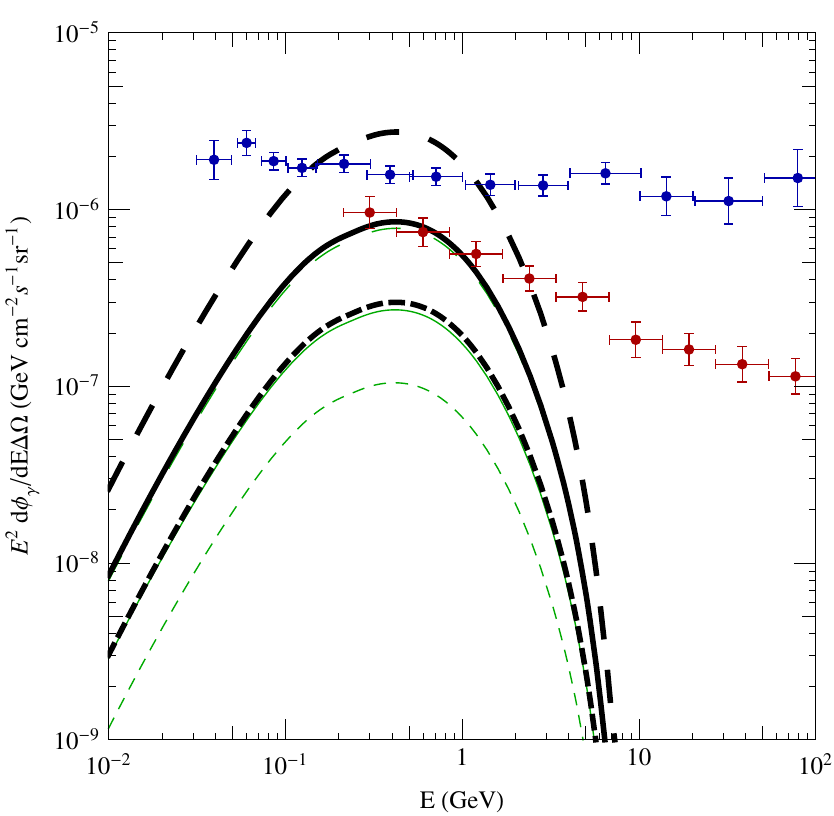}
\end{minipage}
\hspace*{-0.2cm}
\begin{minipage}[t]{0.5\textwidth}
\includegraphics[width=0.9\columnwidth]{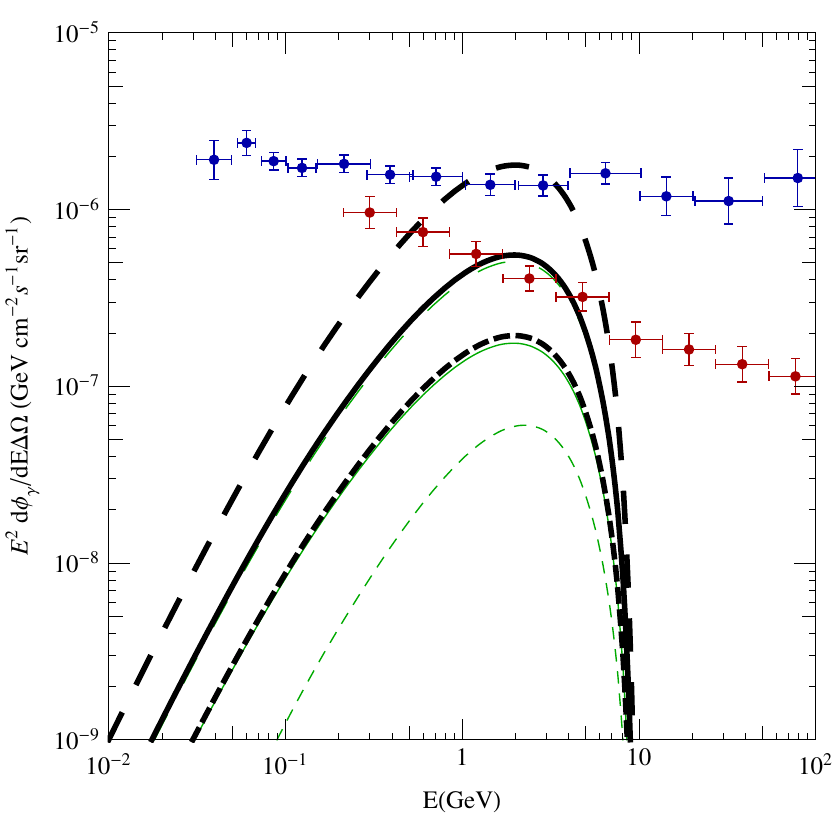}
\end{minipage}
\caption{Diffuse photon emission from a 10 GeV dark matter particle for $BR_{b\bar{b}}=100\%$ on the left, and with annihilation into $BR_{\tau^+\tau^-}=100\%$ on the right. For the $b\bar{b}$ channel $\svann= 2.6 \ 10^{-26} \rm cm^{3} \rm s^{-1}$, while for the $\tau^+\tau^-$ case the thermal cross-section is fixed to $3.3 \ 10^{-26} {\rm cm^{3} s^{-1}}$. The color code is as in Fig.~\ref{fig:boost}. The red points (below) are the measurements of the diffuse emission by Fermi-LAT~\cite{Abdo:2010nz} and the blue points (above) are those by EGRET~\cite{Sreekumar:1997un,Strong:2004ry}.}
\label{fig:e2flux}
\end{figure}

The measured fluxes $\phi_{i}^{\rm IGRB}$ and the errors $\sigma_{i}$ for the diffuse emission are those given by the Fermi-LAT collaboration, listed in Table I of~\cite{Abdo:2010nz} in the energy range from 0.2 GeV up to 102.4 GeV. The measurements arise from a full sky fit, for latitude $b\geq 10^{\circ}$ and from the data of the initial 10 months of the Fermi mission. From the isotropic diffuse component measured by Fermi, the IGRB flux is obtained subtracting the diffuse galactic emission, point sources and a cosmic rays background. The measured IGRB spectrum is compatible with a featureless power law with index $-2.41$~\cite{Abdo:2010nz}. In this light we derive conservative upper bounds on the annihilation cross-section of a WIMP, namely the dark matter signal should not exceed the measured flux in any individual bin $i$ by more than a given amount
\begin{equation}\label{eq:ub}
 \phi_{i}^{\rm th} \leq \phi_{i}^{\rm IGRB}+ n\,  \sigma_{i}\,. 
 \end{equation}
The index $n$ indicates the significance of the upper bound, $n=1.64$ corresponds to the $95\%$ confidence level (C.L.), assuming that the probability distributions for the intensity in each bin are Gaussian and independent from each other. The limits on the DM annihilation rate are derived taking the lowest value of $\svann$ for which Eq.~(\ref{eq:ub}) is not satisfied. In order to explain the Fermi IGRB data one needs to add a background contribution in each energy bin where the DM flux is too small to account for the measured intensity.
\begin{figure}
\begin{minipage}[t]{0.5\textwidth}
\centering
\includegraphics[width=0.9\columnwidth]{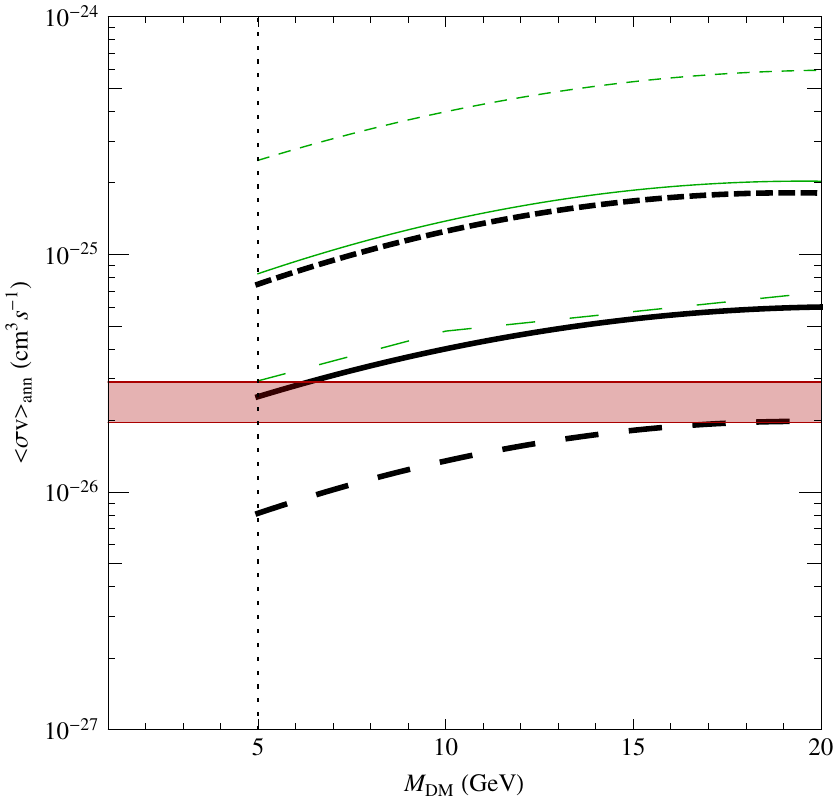}
\end{minipage}
\hspace*{-0.2cm}
\begin{minipage}[t]{0.5\textwidth}
\includegraphics[width=0.9\columnwidth]{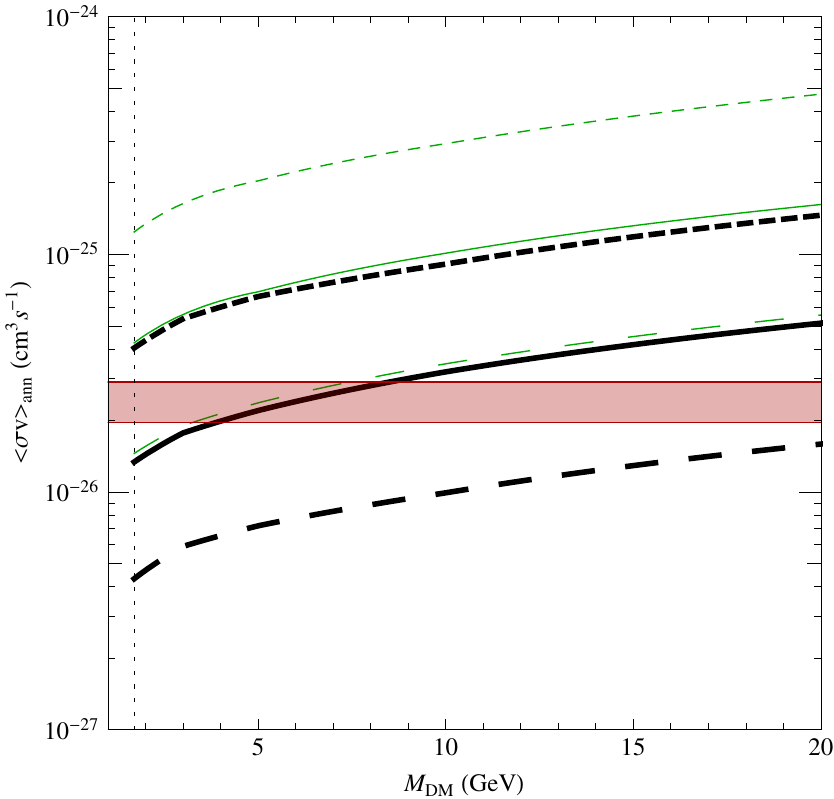}
\end{minipage}
\caption{Fermi-LAT upper bounds in the plane $\svann$ versus the DM mass $M_{DM}$ at $95\%$ C.L. for two different annihilation channels: on the left $BR_{b\bar{b}}=100\%$ and on the right $BR_{\tau^+\tau^-}=100\%$. The horizontal red (gray) region corresponds to the WMAP-7yrs $5-\sigma$ for the thermal annihilation cross-section. {The vertical dotted line indicates the energy threshold for $b, \bar{b}$ and $\tau^{\pm}$ production, on the left and right respectively}. The color code is as in Fig.~\ref{fig:boost}.}
\label{fig:up_mi}
\end{figure}

In Figure~\ref{fig:e2flux}, the predicted IGRB photon flux times $E^2$ as a function of the photon energy is shown for a 10 GeV DM particle that annihilates  with a branching ratio of $100\%$ into $b\bar{b}$ or $\tau^+\tau^-$ together with  the Fermi-LAT data (red bins). For the $b\bar{b}$ model, a $\svann = 2.6 \cdot 10^{-26} {\rm cm}^{3} {\rm s}^{-1}$ is already excluded for $M_{\rm min}=10^{-8} M_{\odot}$ and $c_{vir}=C_{\rm PL}$, while it is fully compatible for $M_{\rm min}=10^{-6} M_{\odot}$ and $c_{vir}=C_{\rm PL}$ or $M_{\rm min}=10^{-8} M_{\odot}$ and $c_{vir}=C_{\rm WMAP}$. From the shape of the flux, we see that the upper bounds will come from the first, second or at most third bin (depending on the dark matter mass) of the Fermi-LAT data, namely from measured photon energies lower than 2 GeV. The $\tau^+\tau^-$ case is analogous, again $M_{\rm min}=10^{-8} M_{\odot}$ and $c_{vir}=C_{\rm PL}$ excludes a  cross-section of $3.3 \cdot 10^{-26} {\rm cm}^{3} {\rm s}^{-1}$, while this value is compatible for all other configurations of the astrophysical parameters. However in this case the most constraining energy bins are from the second to the fifth: the relevant energy range is more close to the photon production threshold, namely the dark matter mass. The contribution to the extra galactic photon flux comes from low redshifts.

Scanning over the mass range from 1 GeV up to 20 GeV, we obtain the upper bounds in the plane $\svann$ {\it vs} the dark matter mass $M_{DM}$, Figure~\ref{fig:up_mi}. The pure $b\bar{b}$ final state case is the left panel, while the right plot is given for an annihilation into $\tau^+\tau^-$. Comparing these two model-independent cases, we note that the leptonic final state give more stringent upper bounds, for the same assumption on the astrophysical parameters, due to the different shape of the photon spectrum. In example, the $\svann$ values that lead to a relic abundance in the WMAP range, are always compatible with the Fermi data for $BR_{b\bar{b}}=100\%$, while they are excluded at $95\%$ C.L. for $BR_{\tau^+\tau^-}=100\%$ and $M_{DM} \leq 7$ GeV, with $M_{\rm min}=10^{-6} M_{\odot}$ and $C_{\rm PL}$. As already noted, the different choice in $M_{\rm min}$ moves up or down the bounds by a factor of $\sim 3$. The bounds from the $b\bar{b}$ are compatible with those in~\cite{Abazajian:2010sq,Abdo:2010dk}. The pure $\tau^+\tau^-$ has not yet been discussed for such low DM masses.

\section{The singlet scalar with Higgs portal and a Dirac fermion with a $Z^\prime$}
\label{sec:SCALAR}
\begin{figure}[t!]
\centering
\includegraphics[width=0.5\columnwidth]{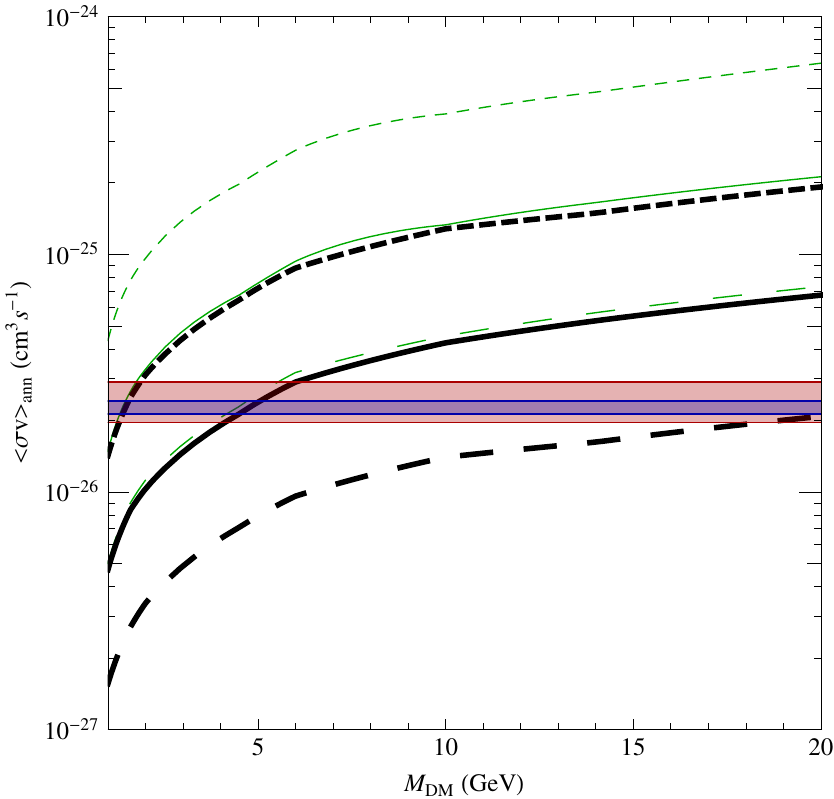}
\caption{Fermi-LAT exclusion limits (colours as in Fig.~\ref{fig:up_mi})  on the singlet dark matter candidate with ``Higgs-like" annihilation channels, together with the $2-\sigma$ (purple, dark grey) and $5-\sigma$ (pink, light grey) WMAP-7yrs regions.}
\label{fig:singlet}
\end{figure}

One may say that they are essentially two classes of models of dark matter. To the first class belong models within which dark matter is a by-product, the secondary effect of a deeper goal. Such are the neutralino and the axion, and, to some extent, DM candidates in theories with extra dimensions (for a review see~\cite{Bertone:2004pz}). The other class encompasses the many models which start with the DM problem, sometime in a minimalist way, sometime with a broader scope in mind. The first class may be nobler, but, with the advent of new data, the bottom-up approach to DM has gained some momentum. 
In particular, the interpretation of the CoGeNT or DAMA data as being due to DM has motivated much works, both on the possibility to explain the data with a light neutralino~\cite{Bottino:2003iu,Dudas:2008eq,Kuflik:2010ah,Bottino:2009km,Das:2010ww,Bae:2010hr,Feldman:2010ke} and on new models (see {\em e.g.}~\cite{Foot:2008nw,Chang:2008xa,Khlopov:2008ki,Bernabei:2008mv,Feng:2008dz,Masso:2009mu,Petriello:2008jj,Savage:2008er,Chang:2008gd,Cui:2009xq,Bandyopadhyay:2010cc,Cohen:2010kn,Kuflik:2010ah,Das:2010ww,Bae:2010hr,Barger:2010yn,Farzan:2010mr,Mambrini:2010dq}). In this section we consider two very simple light WIMP candidates. These models have in common the facts that  they are based on minimal extensions of the SM and yet may explain the direct detection data while having a relic abundance that agrees with cosmological observations. The first model is based on a real scalar singlet, which is interacting through the Higgs portal with the SM degrees of freedom~\cite{McDonald:1993ex,Burgess:2000yq,Andreas:2008xy}. The second model invokes a singlet Dirac fermion, interacting through the kinetic mixing portal, {\em i.e.} via a new $Z^\prime$ gauge boson~\cite{Mambrini:2010dq,Cheung:2010az,Dudas:2009uq}. 

That a real scalar singlet interacting through the Higgs may be in agreement with both CoGeNT-DAMA and WMAP has been argued in~\cite{Andreas:2008xy,Arina:2009um,Andreas:2010dz}. This is implicit in other works. In particular in the results of~\cite{Fitzpatrick:2010em}, which are based on an effective operator approach (see in particular Figure 4, lower left plot). We may also refer to~\cite{Barger:2010yn}, which is based on a singlet complex scalar, in particular Figure 7. In practice, these models differ only in the choice of couplings to the SM degrees of freedom, which in the case of~\cite{Andreas:2008xy,Andreas:2010dz} are simply proportional to the SM fermions mass (``Higgs-like'' couplings) and thus more constrained. These scalar DM models also have in common the fact that, at the low energies relevant for the CoGeNT-DAMA regions and the calculation of the relic abundance from freeze-out, the ratio of the annihilation and elastic scattering on nuclei cross-sections only depends on the DM mass. In particular, for the case of one real scalar singlet S of mass $m_{S}$ interacting through the Higgs, the ratio is 
\begin{equation}
\label{eq:ratio}
\sum_f \frac{ \sigma(S S \rightarrow \bar{f} f) v_{rel}}{\sigma(S N \rightarrow  S N)}= \sum_f
\frac{n_c m^2_f}{f^2 m^2_N \mu^2_n}\frac{(m_{S}^2-m_f^2)^{3/2}}{m_{S}}\,,
\end{equation}
where $n_c =3(1)$ for quarks (leptons), $\mu_n$ is the DM-nucleon reduced mass and the factor $f$ parameterizes the Higgs to nucleons coupling from the trace anomaly, which has central value $f\sim 0.3$ \cite{Andreas:2008xy}. Taking into account the branching ratio for annihilation of a light S into SM fermions (see Table II), the limits on the annihilation cross-section of the S from the IGRB measured by Fermi-LAT are shown in Figure~\ref{fig:singlet}. Which limiting curve we may believe is most relevant depends strongly on the minimum halo mass $M_{\rm min}$. An interesting feature is that, after freeze-out, a particle with ``Higgs-like'' couplings to SM degrees of freedom, has very little interactions with particles from the thermal bath. Indeed we may neglect the Yukawa coupling with the electron and positrons, while the protons and neutrons are very few in the thermal bath. Using Eq.~(\ref{eq:loeb})
 we  estimated  $M_{\rm min} \sim 10^{-8} \,M_\odot$. To this parameter, together with $C_{\rm PL}$, corresponds the strongest constraint we may get based on the IGRB (long dashed curves in Figure~\ref{fig:singlet}). In this plot we see that all the $\svann$ values that account for the WMAP abundance are excluded at $95\%$ C.L., while more relaxed constraints come from either $M_{\rm min}=10^{-6} M_{\odot}$ and $C_{\rm PL}$ (solid black thick curve) or $M_{\rm min}=10^{-8} M_{\odot}$ and $C_{\rm WMAP}$ (long-dashed thin green line). Annihilation cross-sections of the order $3 \cdot 10^{-26}{\rm cm^3 s^{-1}}$ are excluded only for very light scalar masses, $m_S \leq 7$ GeV. 
\begin{table}[t]
  \begin{center}
    \begin{tabular}{c|c|c|c|c}
& \multicolumn{4}{|c}{Branching ratios}   \\
       \hline 
       $m_S$  & $b\bar b$ & $c\bar c$ & $\tau^+\tau^-$ & others  \\
      \hline\hline
      20 GeV &  85 \% & 5 \% & 9 \% & $\sim$ 1 \%\\
      10 GeV & 83 \%  & 7\% & 10 \% & $\lsim $ 1 \%\\
       5 GeV & 16 \% & 36 \% & 42 \% & $\sim$ 5 \%\\ 
     2 GeV & $\backslash$  & 69 \% & 22 \% & $\sim$ 9 \% \\
      \hline\hline
    \end{tabular}    
       \label{tab:dwarfs}
  \end{center}
\caption{Branching ratios in the main annihilation channels of the scalar singlet, for various light candidates.}
\end{table}
\begin{figure}[t!]
\centering
\includegraphics[width=0.5\columnwidth]{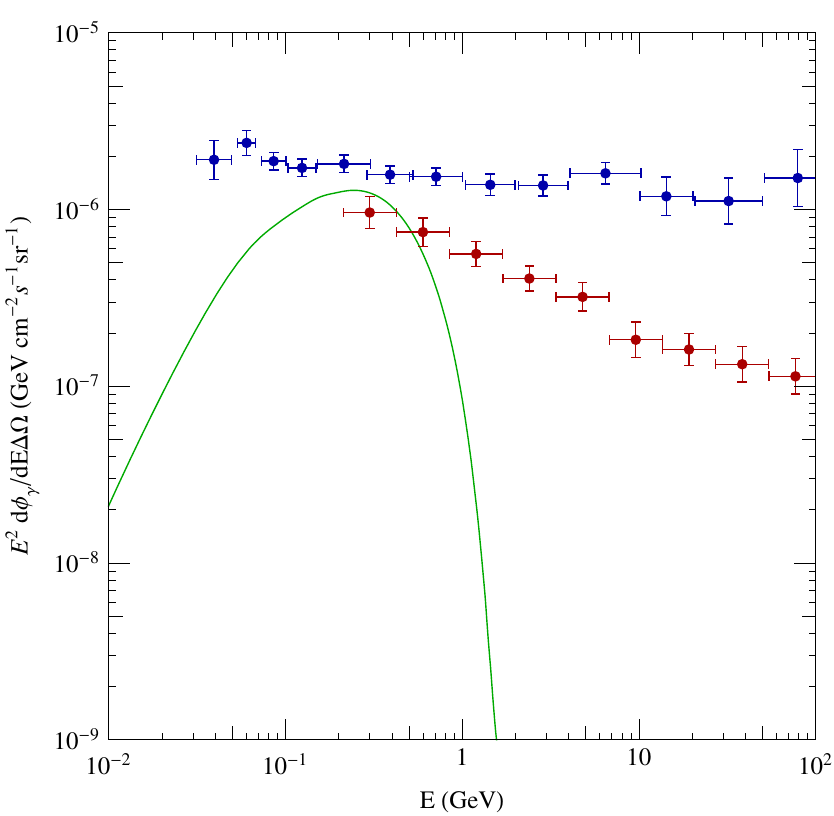}
\caption{Diffuse photon emission for a $m_S = 2$ GeV candidate, the lightest DM particle we consider here,  compared to the Fermi-LAT (in red, below) and EGRET data (in blue, above). Because the EGRET data points are above those from Fermi-LAT, only the latter is constraining for the mass range relevant for the model. From Figure~\ref{fig:singlet} we used the maximum allowed $\svann = 3.2 \cdot 10^{-26} {\rm cm}^{3} {\rm s}^{-1}$ and the color code as in Figure~\ref{fig:e2flux}.}
\label{fig:singletSI_EGRET}
\end{figure}

\begin{figure}[t!]
\begin{minipage}[t]{0.5\textwidth}
\centering
\includegraphics[width=0.9\columnwidth]{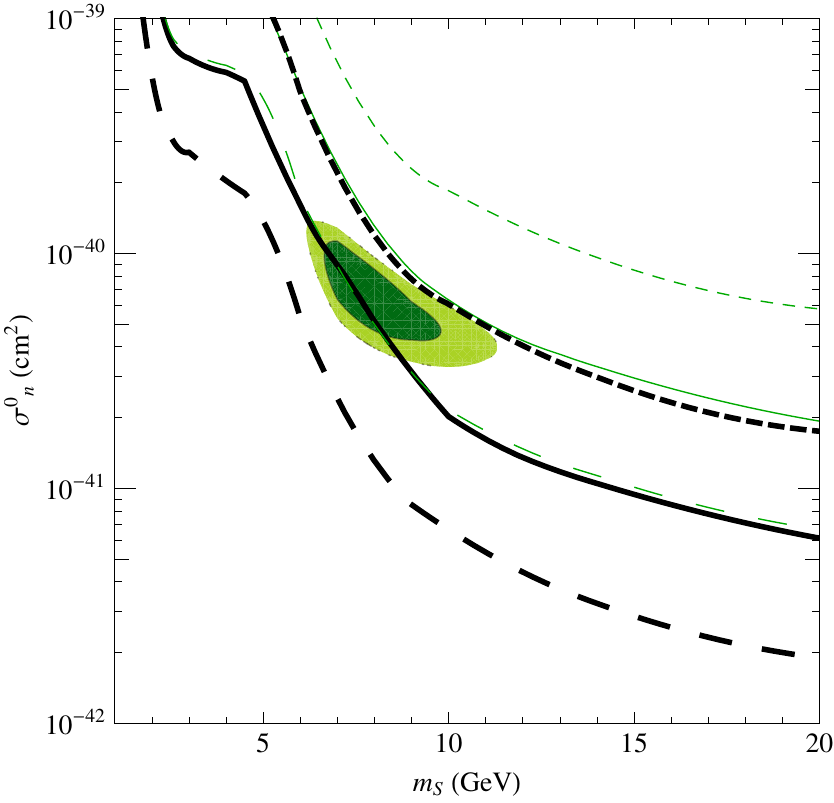}
\end{minipage}
\hspace*{-0.2cm}
\begin{minipage}[t]{0.5\textwidth}
\centering
\includegraphics[width=0.9\columnwidth]{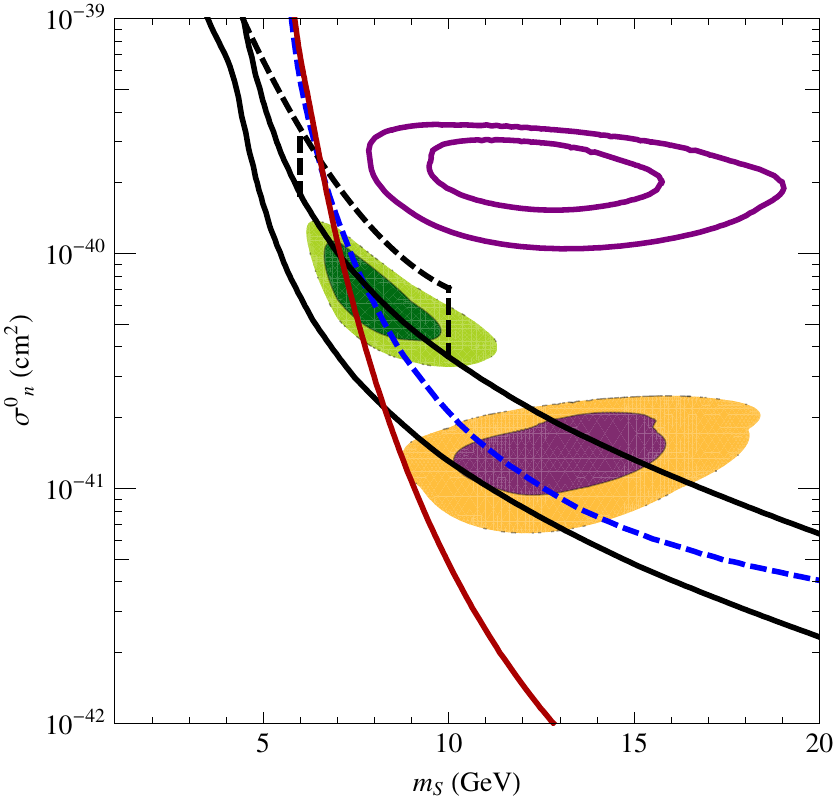}
\end{minipage}
\caption{ The SI cross-section ($\sigma^0_{n}$) {\em vs} the scalar singlet mass ($m_S$).
On the left, the region (in green) corresponds to CoGeNT~\cite{Aalseth:2010vx} (minimum $\chi^2$, with contours at 90 and 99.9\% C.L.). The black thick and green thin lines are as in Fig.~\ref{fig:boost} and denote the exclusion limits from the Fermi-LAT diffuse gamma-rays flux at $95\%$ C.L.. On the right, we give, for the sake of comparison, the corresponding exclusion limits from direct detection experiments. We include the DAMA regions~\cite{Bernabei:2008yi} (goodness-of-fit, also at 90 and 99.9\% C.L.) both with (below CoGeNT, purple/orange) and without (above CoGeNT, purple, no fill) channelling. The blue (short-dashed) line is the 90\% C.L. exclusion limit from CDMS-Si~\cite{Akerib:2005kh}. The red solid curve is the $90\%$ exclusion limit from Xenon100~\cite{Aprile:2010um}, using LeffMin and a threshold at 3 PhotoElectrons. For more details on the experimental upper bounds and conventions see~\cite{Andreas:2010dz}.  On the right, we include the envelope of $\sigma^0_{n}-m_S$ parameters consistent with WMAP. The continuous region corresponds to the standard assumption of a QCD phase transition at $T_c = 150$ MeV. The black dashed lines extend this domain for other, possible but less likely, values of $T_c$, from left to right $T_c = 300$ MeV and $T_c=500$ MeV respectively.}
\label{fig:singletSI}
\end{figure}
Since the lowest energy bin measured by Fermi-LAT is most relevant for light WIMPs, one may ask whether the EGRET data, which extend to lower energies could give further constraints~\cite{Sreekumar:1997un,Strong:2004ry}. However, this is not the case for the range of mass we consider,  as is shown in Figure~\ref{fig:singletSI_EGRET} for a $m_S = 2$ GeV candidate. A similar conclusion holds for the Dirac fermion candidate. Of course, this does not mean that EGRET data are not relevant for even lighter DM candidates, $M_{DM}\lsim 1$ GeV, but only that such candidates are beyond the scope of the models discussed here. 

Using the Eq.~(\ref{eq:ratio}), we may transpose the limits on the annihilation cross-section into limits on the elastic scattering cross-section  on nuclei, and plot them in the $\sigma_{n}^0-m_{S}$ plane, Figure~\ref{fig:singletSI} on the left, together with the  CoGeNT-DAMA regions and the exclusion limits set by CDMS-Si and Xenon100, which were  computed as in~\cite{Andreas:2010dz}. Following the same convention as in the other figures,  the upper bounds from the IGRB are shown in green or black. For $M_{\rm min}=10^{-8} M_{\odot}$ and a power law concentration parameter the DAMA and CoGeNT regions are both excluded at $95\%$ C.L., while the same minimum halo mass with $C_{\rm WMAP}$ gives limits which are marginally compatible with the CoGeNT region. A choice of $M_{\rm min}=10^{-6} M_{\odot}$ and $C_{\rm WMAP}$ is totally compatible with the CoGeNT region and can be accommodated with the DAMA region (see footnote 7).   

For  comparison, we also give in Figure~\ref{fig:singletSI} (right panel) the region of the  $\sigma_{n}^0-m_S$ plane consistent with WMAP.  In this figure, the black continuous lines is based on the hypothesis that the QCD phase transition occurred at a temperature $T_c = 150$ MeV (which is the standard assumption both for DarkSusy~\cite{Gondolo:2004sc} and Micromegas~\cite{Belanger:2006is}). Freeze-out after the QCD phase transition would require an annihilation cross-section which is about twice that for a standard WIMP, $\svann \sim 3 \cdot 10^{-26}$ cm$^3$s$^{-1}$ (see for instance~\cite{Bottino:2003iu}). Since typically $x_{fo} = M_{DM}/T_{fo} \sim 20$, this effect is only relevant for $M_{DM} \lsim 3$ GeV. However, we may contemplate the possibility that the QCD phase transition took place at a higher temperature, for instance $150$ MeV $<T_c < 500$ MeV, which is relevant for a candidate $M_{DM} \lsim 10$ GeV, and thus for the CoGeNT-DAMA regions.\footnote{This effect may be relevant in the light of the many possible uncertainties that may be hidden in a Figure like~\ref{fig:singletSI} (plot on the right). For instance, in~\cite{Savage:2010tg}, it is argued that channelling may, after all, not be very relevant for the interpretation of the DAMA data, so the lower of the two regions corresponding to DAMA most presumably does not exists. There is however still substantial freedom just on the experimental side, and, based on the current uncertainties on parameters like quenching, the authors of~\cite{Hooper:2010uy} argued quite convincingly that, not only, both the DAMA (without channelling) and CoGeNT may be consistent with each others but, moreover, are not  excluded by the current exclusion limits.}
\begin{figure}[t!]
\centering
\includegraphics[width=0.5\columnwidth]{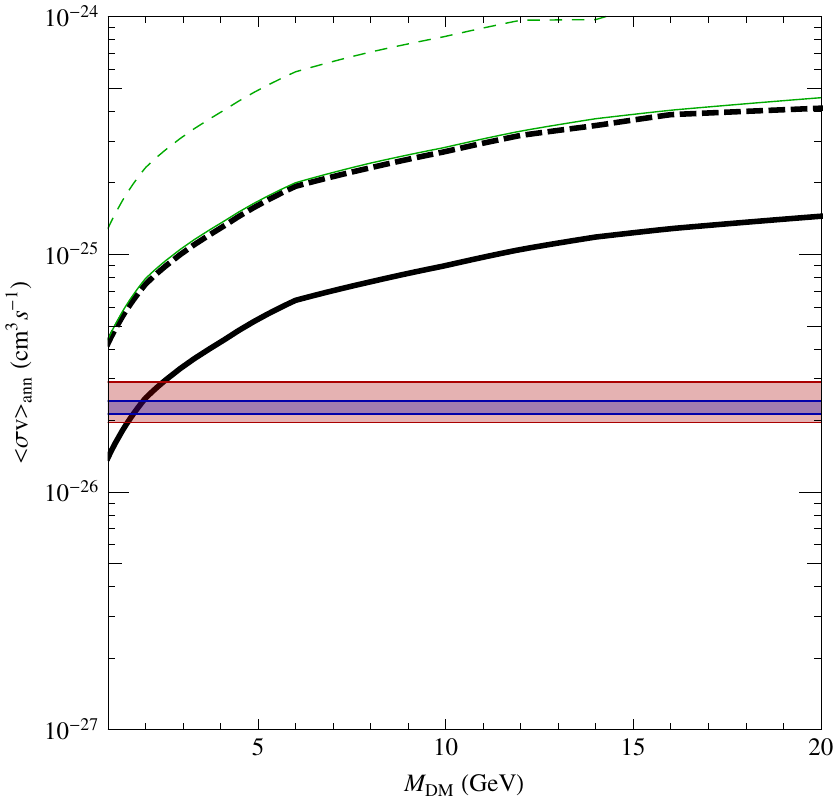}
\caption{The same as in Fig.~\ref{fig:singlet} for the singlet Dirac fermion candidate, through annihilation channels with 
``Z-like" couplings.}
\label{fig:psiSI}
\end{figure}

One may notice  that the constraints from IGRB extend to candidates which are even (slightly) lighter than those of interest for CoGeNT-DAMA, and which are not constrained by direct detection experiments (so far). In particular, given the conservative constraint corresponding to the continuous black curve in the left panel of Figure~\ref{fig:singletSI}, we may conclude that candidates lighter than 5 GeV, and consistent with WMAP, are excluded at 95 \% C.L. by the Fermi-LAT data on the IGRB.

For the sake of comparison with the singlet scalar candidate,  we consider a model that consists of a singlet Dirac fermion candidate $\psi$, charged under a broken $U(1)^\prime$ gauge group, which may interact with the SM degrees of freedom through the kinetic mixing portal. In~\cite{Mambrini:2010dq}, it has been shown that such a candidate may be consistent with both WMAP and the CoGeNT-DAMA regions as well as constraints from LEPI on the $Z$ invisible width. We notice that this result is {\em a priori} in contradiction with the one drawn in~\cite{Andreas:2008xy} and~\cite{Fitzpatrick:2010em}, where it has been shown that a singlet Dirac DM candidate (for a single, vector or  scalar annihilation channel) that fits the CoGeNT-DAMA regions has  a too large relic abundance.  However, there is no magic, as the conclusion in~\cite{Andreas:2008xy,Fitzpatrick:2010em} has been reached by assuming that there exists a one-to-one correspondence between the annihilation and scattering cross-section (for fixed dark matter mass), while in~\cite{Mambrini:2010dq} the proximity of the $Z^\prime$ pole is used to enhance this annihilation cross-section so as to get the right relic abundance. Incidentally, this will prevent us from presenting the exclusion limits in the $\sigma_n^0-M_{DM}$ plane for this model, for the ratio of $\svann$ and $\sigma_n^0$ depends on $m_{Z^\prime}$ and $\Gamma_{Z^\prime}$, 
\begin{equation}
\label{eq:Zprime}
\sum_f 
\frac{ \sigma( \bar{\psi}\psi \rightarrow \bar{f} f) v_{rel}}{\sigma(\psi N \rightarrow  \psi N)}  
= \sum_f \frac{n_c}{4}
\frac{\Big(1-\frac{m^2_f}{m^2_\psi}\Big)^{1/2}}{\mu^2_n} \Big( (1-
\frac{4 m^2_\psi}{m^2_{Z^\prime}})^2
+\frac{\Gamma_{Z^\prime}^2}{m^2_{Z^\prime}}\Big)^{-1} \Big(2 m^2_\psi
(v^2_f+a^2_f)+(v^2_f-a^2_f) m^2_f\Big)\,.
\end{equation} 
In Eq.~(\ref{eq:Zprime}), $a_f$ and $v_f$ are the SM model axial and vector couplings of the fermions to the $Z$. The limits on the annihilation cross-section of $\psi$ from the IGRB are shown in Figure~\ref{fig:psiSI}. That the $Z^\prime$ has ``Z-like'' couplings to the SM degrees of freedom (compared to the ``Higgs-like'' couplings of the scalar singlet) has two interesting consequences regarding the constraints from the 
 IGRB. First the annihilation channels are distinct, with a spectrum that is harder with respect to the scalar singlet one, and lead to the constraints from Fermi-LAT to extend to rather light DM candidates. Also, because of the abundance of the $e^+/e^-$ (and neutrinos) in the thermal bath, the kinetic decoupling (compared to the chemical freeze-out that fixes the relic abundance) of a DM candidate with ``Z-like" couplings  occurs at a much smaller temperature, which, following~\cite{Green:2005fa}, we estimated to be ${\cal O}(10\  \mbox{\rm MeV})$. This implies that the minimum halo mass that may form in the model is substantially larger for the Dirac fermion than in the case of the singlet scalar,  $M_{\rm min} \sim 10^{-4} M_\odot$ from  Eqs.~(\ref{eq:minhalomass},\ref{eq:loeb}). This is the reason why we only plot the exclusion limits for $M_{\rm min} \sim 10^{-6} M_\odot$  and $M_{\rm min} \sim 10^{-4} M_\odot$ in Figure~\ref{fig:psiSI}. In this case the most stringent limit, given by $M_{\rm min}=10^{-6} M_\odot$ and $C_{\rm PL}$ excludes at $95\%$ C.L. the WMAP region for $m_{\psi} < 5$ GeV. All the other curves only marginally affect $\svann \sim 3 \cdot 10^{-26} {\rm cm^3 s^{-1}}$. 
 
Altogether, we may conclude that, although they are both viable candidates with respect to WMAP and CoGeNT-DAMA, the constraints from the IGRB are substantially weaker for the $\psi$ candidate, with ``Z-like'' couplings, than for the $S$ candidate with ``Higgs-like'' couplings, a feature we believe should be of interest for  model building.

Needless to say, such constraints should be taken with a grain of salt, as they suffer from the many astrophysical uncertainties, regarding for instance the profile of dark matter in halos, and other parameters which enter in the estimation of the extragalactic boost (even the use of the Press-Schechter formula may be a matter of discussion, as is the extrapolation of over many decades of the known power spectrum of the DM inhomogeneities). Regarding the profile we have chosen use the somewhat standard NFW choice, like in~\cite{Abdo:2010dk,Hutsi:2010ai}; clearly a different choice, like the Einasto or Burkert profiles would give less stringent constraints.

\section{Conclusions and prospects}
\label{sec:conclusions}

In this work, we have studied further the constraints that may be set, using the Fermi-LAT data on the isotropic diffuse gamma-ray emission, on the parameter space of two simple, albeit generic, DM candidates, which has been shown to be consistent both with the WMAP relic abundance, and the regions favoured by CoGeNT-DAMA~\cite{Andreas:2010dz,Mambrini:2010dq}. The constraints we get are consistent with results which may be inferred from other current works on constraints on DM from extragalactic gamma-ray fluxes~\cite{Abdo:2010dk,Abazajian:2010sq}. We have however extended them to lower DM masses and make explicitly the connection with the CoGeNT-DAMA results tentatively interpreted as being due to the elastic scattering of DM particles. Our main results are summarized in Figure~\ref{fig:singlet} for the case of a singlet scalar candidate $S$ interacting through the Higgs portal (``Higgs-like'' couplings to SM degrees of freedom) and Figure~\ref{fig:psiSI} for the case of Dirac fermion candidate $\psi$ interacting through a $Z^\prime$ (``Z-like'' couplings). In the former case, we have used  the one-to-one correspondence between the elastic cross-section and the annihilation cross-section, Eq.~(\ref{eq:ratio}) to express the limit from the isotropic gamma-ray background radiation (IGRB) directly in the $\sigma_n^0-m_S$ plane, Figure~\ref{fig:singletSI}, which is the most direct way to express the constraints on  CoGeNT-DAMA from the IGRB. This works only for the scalar candidate, because it has so few free parameters, but would also apply to an approach based on effective operators. 

Like all indirect signatures, the constraints on DM derived from the data on the IGRB are quite sensitive to astrophysical uncertainties. As such, they are perhaps best thought as being complementary to those that could be extracted from other observations. Most relevant are the constraints based on gamma-rays from dwarf spheroidal galaxies (dSphs), also observed by Fermi-LAT, which may also exclude the light WIMPs, $M_{DM} \lsim 10$ GeV, assuming a NFW profile of DM distribution~\cite{Fitzpatrick:2010em,Andreas:2010dz,GardeIDM2010}. Unlike the constraints from the IGRB, the constraints from dSphs are the same for the scalar singlet and the Dirac singlet, as they both have S-wave, unsuppressed,  annihilation into SM particles. A potentially important indirect constraints on a light WIMP could come from measurements of the synchrotron radiation near the galactic centre, and in particular from galaxy clusters~\cite{Regis:2008ij,Hooper:2008zg,Borriello:2008gy,Crocker:2010gy,Boehm:2002yz}.  Such constraints are {\em a priori} the strongest for candidates that annihilate directly into $e^+ e^-$ pairs (see {\em e.g.}~\cite{Boehm:2002yz}), and so we  expect that they may be stronger for the Dirac singlet interacting through a $Z^\prime$ than for the singlet scalar candidate discussed here. A multi-wavelengths analysis of the light WIMPs mass range could thus be of interest, but this is beyond the scope of the present work.

\section*{Acknowledgments}
One of us (M.T.) would like to thank Y. Mambrini for useful discussions. We also thank M. Gustafsson for  useful comments.  
Our work is supported by the FNRS-FRS, 
the IISN and the Belgian Science Policy (IAP VI-11).

\bibliography{bibliography_chiara}

\end{document}